\RequirePackage{lineno} 
\documentclass[prc,aps,floatfix,showpacs,showkeys,twocolumn,superscriptaddress,letterpaper,10pt]{revtex4-1}

\usepackage{color}
\usepackage[normalem]{ulem}
\input epsf

\usepackage{amsmath}
\usepackage{graphicx}
\usepackage{url}
\usepackage{color}
\usepackage{subfigure}
\usepackage{longtable}
\usepackage{dcolumn}

\newcommand{\GPDHT}{\langle H_T \rangle}
\newcommand{\GPDETbar}{\langle \bar{E}_T \rangle}



\begin{document}

\setpagewiselinenumbers


\title{ Exclusive $\eta$  electroproduction at $W>2$ GeV with CLAS\\ and transversity generalized parton distributions}

\newcommand*{\ANL}{Argonne National Laboratory, Argonne, Illinois 60439}
\newcommand*{\ANLindex}{1}
\affiliation{\ANL}
\newcommand*{\ASU}{Arizona State University, Tempe, Arizona 85287-1504}
\newcommand*{\ASUindex}{2}
\affiliation{\ASU}
\newcommand*{\CSUDH}{California State University, Dominguez Hills, Carson, CA 90747}
\newcommand*{\CSUDHindex}{3}
\affiliation{\CSUDH}
\newcommand*{\CANISIUS}{Canisius College, Buffalo, NY}
\newcommand*{\CANISIUSindex}{4}
\affiliation{\CANISIUS}
\newcommand*{\CMU}{Carnegie Mellon University, Pittsburgh, Pennsylvania 15213}
\newcommand*{\CMUindex}{5}
\affiliation{\CMU}
\newcommand*{\CUA}{Catholic University of America, Washington, D.C. 20064}
\newcommand*{\CUAindex}{6}
\affiliation{\CUA}
\newcommand*{\SACLAY}{Irfu/SPhN, CEA, Universit\'e Paris-Saclay, 91191 Gif-sur-Yvette, France}
\newcommand*{\SACLAYindex}{7}
\affiliation{\SACLAY}
\newcommand*{\CNU}{Christopher Newport University, Newport News, Virginia 23606}
\newcommand*{\CNUindex}{8}
\affiliation{\CNU}
\newcommand*{\UCONN}{University of Connecticut, Storrs, Connecticut 06269}
\newcommand*{\UCONNindex}{9}
\affiliation{\UCONN}
\newcommand*{\FU}{Fairfield University, Fairfield CT 06824}
\newcommand*{\FUindex}{10}
\affiliation{\FU}
\newcommand*{\FIU}{Florida International University, Miami, Florida 33199}
\newcommand*{\FIUindex}{11}
\affiliation{\FIU}
\newcommand*{\FSU}{Florida State University, Tallahassee, Florida 32306}
\newcommand*{\FSUindex}{12}
\affiliation{\FSU}
\newcommand*{\Genova}{Universit$\grave{a}$ di Genova, 16146 Genova, Italy}
\newcommand*{\Genovaindex}{12}
\affiliation{\Genova}
\newcommand*{\GWUI}{The George Washington University, Washington, DC 20052}
\newcommand*{\GWUIindex}{13}
\affiliation{\GWUI}
\newcommand*{\ISU}{Idaho State University, Pocatello, Idaho 83209}
\newcommand*{\ISUindex}{14}
\affiliation{\ISU}
\newcommand*{\INFNFE}{INFN, Sezione di Ferrara, 44100 Ferrara, Italy}
\newcommand*{\INFNFEindex}{15}
\affiliation{\INFNFE}
\newcommand*{\INFNFR}{INFN, Laboratori Nazionali di Frascati, 00044 Frascati, Italy}
\newcommand*{\INFNFRindex}{16}
\affiliation{\INFNFR}
\newcommand*{\INFNGE}{INFN, Sezione di Genova, 16146 Genova, Italy}
\newcommand*{\INFNGEindex}{17}
\affiliation{\INFNGE}
\newcommand*{\INFNRO}{INFN, Sezione di Roma Tor Vergata, 00133 Rome, Italy}
\newcommand*{\INFNROindex}{18}
\affiliation{\INFNRO}
\newcommand*{\INFNTUR}{INFN, Sezione di Torino, 10125 Torino, Italy}
\newcommand*{\INFNTURindex}{19}
\affiliation{\INFNTUR}
\newcommand*{\ORSAY}{Institut de Physique Nucl\'eaire, CNRS/IN2P3 and Universit\'e Paris Sud, Orsay, France}
\newcommand*{\ORSAYindex}{20}
\affiliation{\ORSAY}
\newcommand*{\ITEP}{Institute of Theoretical and Experimental Physics, Moscow, 117218, Russia}
\newcommand*{\ITEPindex}{21}
\affiliation{\ITEP}
\newcommand*{\JMU}{James Madison University, Harrisonburg, Virginia 22807}
\newcommand*{\JMUindex}{22}
\affiliation{\JMU}
\newcommand*{\KNU}{Kyungpook National University, Daegu 702-701, Republic of Korea}
\newcommand*{\KNUindex}{23}
\affiliation{\KNU}
\newcommand*{\MISS}{Mississippi State University, Mississippi State, MS 39762-5167}
\newcommand*{\MISSindex}{24}
\affiliation{\MISS}
\newcommand*{\UNH}{University of New Hampshire, Durham, New Hampshire 03824-3568}
\newcommand*{\UNHindex}{25}
\affiliation{\UNH}
\newcommand*{\NSU}{Norfolk State University, Norfolk, Virginia 23504}
\newcommand*{\NSUindex}{26}
\affiliation{\NSU}
\newcommand*{\OHIOU}{Ohio University, Athens, Ohio  45701}
\newcommand*{\OHIOUindex}{27}
\affiliation{\OHIOU}
\newcommand*{\ODU}{Old Dominion University, Norfolk, Virginia 23529}
\newcommand*{\ODUindex}{28}
\affiliation{\ODU}
\newcommand*{\URICH}{University of Richmond, Richmond, Virginia 23173}
\newcommand*{\URICHindex}{29}
\affiliation{\URICH}
\newcommand*{\RPI}{Rensselaer Polytechnic Institute, Troy, New York 12180-3590}
\newcommand*{\RPIindex}{29}
\affiliation{\RPI}
\newcommand*{\ROMAII}{Universita' di Roma Tor Vergata, 00133 Rome Italy}
\newcommand*{\ROMAIIindex}{30}
\affiliation{\ROMAII}
\newcommand*{\MSU}{Skobeltsyn Institute of Nuclear Physics, Lomonosov Moscow State University, 119234 Moscow, Russia}
\newcommand*{\MSUindex}{31}
\affiliation{\MSU}
\newcommand*{\SCAROLINA}{University of South Carolina, Columbia, South Carolina 29208}
\newcommand*{\SCAROLINAindex}{32}
\affiliation{\SCAROLINA}
\newcommand*{\TEMPLE}{Temple University,  Philadelphia, PA 19122 }
\newcommand*{\TEMPLEindex}{33}
\affiliation{\TEMPLE}
\newcommand*{\JLAB}{Thomas Jefferson National Accelerator Facility, Newport News, Virginia 23606}
\newcommand*{\JLABindex}{34}
\affiliation{\JLAB}
\newcommand*{\UTFSM}{Universidad T\'{e}cnica Federico Santa Mar\'{i}a, Casilla 110-V Valpara\'{i}so, Chile}
\newcommand*{\UTFSMindex}{35}
\affiliation{\UTFSM}
\newcommand*{\EDINBURGH}{Edinburgh University, Edinburgh EH9 3JZ, United Kingdom}
\newcommand*{\EDINBURGHindex}{36}
\affiliation{\EDINBURGH}
\newcommand*{\GLASGOW}{University of Glasgow, Glasgow G12 8QQ, United Kingdom}
\newcommand*{\GLASGOWindex}{37}
\affiliation{\GLASGOW}
\newcommand*{\VT}{Virginia Tech, Blacksburg, Virginia   24061-0435}
\newcommand*{\VTindex}{38}
\affiliation{\VT}
\newcommand*{\VIRGINIA}{University of Virginia, Charlottesville, Virginia 22901}
\newcommand*{\VIRGINIAindex}{39}
\affiliation{\VIRGINIA}
\newcommand*{\WM}{College of William and Mary, Williamsburg, Virginia 23187-8795}
\newcommand*{\WMindex}{40}
\affiliation{\WM}
\newcommand*{\YEREVAN}{Yerevan Physics Institute, 375036 Yerevan, Armenia}
\newcommand*{\YEREVANindex}{41}
\affiliation{\YEREVAN}
 \newcommand*{\NOWINFNGE}{INFN, Sezione di Genova, 16146 Genova, Italy}

\author {I.~Bedlinskiy}
\affiliation{\ITEP}
\author {V.~Kubarovsky} 
\affiliation{\JLAB}
\affiliation{\RPI}
\author {P.~Stoler} 
\affiliation{\RPI}
\author {K.P. ~Adhikari} 
\affiliation{\MISS}
\author {Z.~Akbar} 
\affiliation{\FSU}
\author {S. ~Anefalos~Pereira} 
\affiliation{\INFNFR}
\author {H.~Avakian} 
\affiliation{\JLAB}
\author {J.~Ball} 
\affiliation{\SACLAY}
\author {N.A.~Baltzell} 
\affiliation{\JLAB}
\affiliation{\SCAROLINA}
\author {M.~Battaglieri} 
\affiliation{\INFNGE}
% ...

\collaboration{The CLAS Collaboration}
\noaffiliation

\date{\today}

\begin{abstract}

The cross section of the exclusive  $\eta$ electroproduction reaction $ep\to e^\prime p^\prime \eta$ was measured at Jefferson Lab  with a 5.75-GeV electron beam and the CLAS detector. Differential cross sections $d^4\sigma/dtdQ^2dx_Bd\phi_\eta$ and  structure functions $\sigma_U = \sigma_T+\epsilon\sigma_L, \sigma_{TT}$ and $\sigma_{LT}$, as functions of $t$ were obtained over a wide range of $Q^2$ and $x_B$.  The $\eta$ structure functions are compared with those previously measured for $\pi^0$ at the same kinematics. 
At low $t$, both  $\pi^0$ and $\eta$ are described reasonably well by generalized parton distributions (GPDs)  in which chiral-odd transversity GPDs are dominant. 
The $\pi^0$ and $\eta$ data, when taken together, can facilitate the flavor decomposition of the transversity GPDs.
\end{abstract}

\maketitle

\section{Introduction}	

\begin{figure}[h]
\includegraphics[width=\columnwidth]{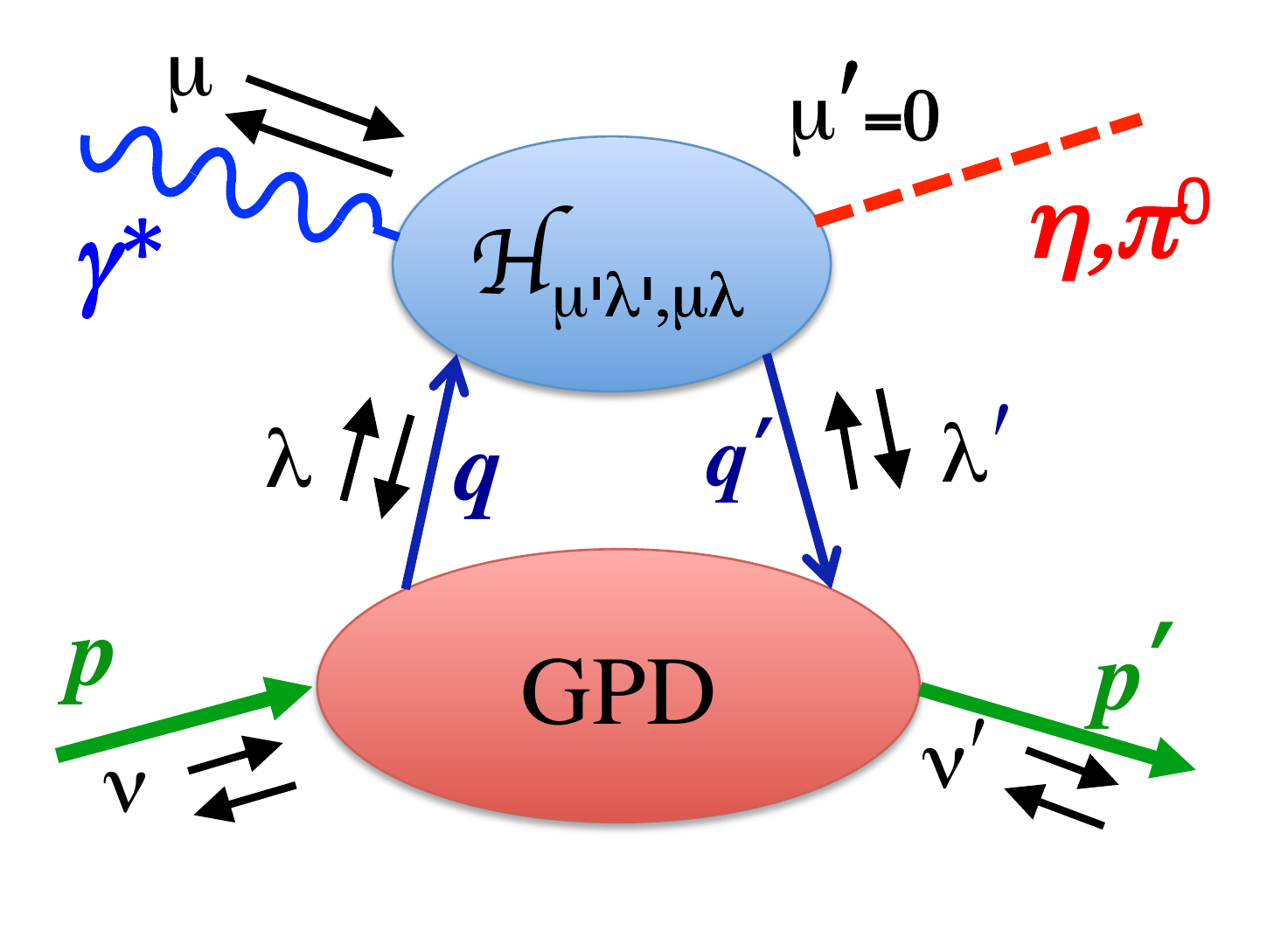}
\caption{
\label{fig:handbag-pi0} 
The handbag diagram for  deeply virtual  $\eta$ and $\pi^0$ production.
The helicities of the initial and final nucleons are denoted by $\nu$ and $\nu^\prime$, of the incident photon and produced meson by $\mu$ and $\mu^\prime$ and of the active initial and final quark by $\lambda$ and $\lambda^\prime$. The arrows in the figure represent schematically the corresponding positive and negative helicities, respectively. 
For final-state pseudoscalar mesons  $\mu^\prime =  0$.
}
\end{figure}

Understanding nucleon structure in terms of the fundamental degrees of freedom of Quantum Chromodynamics (QCD) is one of the main goals in the theory of strong interactions.   Exclusive reactions may provide information about the quark and gluon distributions encoded in  Generalized Parton Distributions (GPDs), which are accessed via application of the handbag mechanism  
\cite{Ji:1996ek,*Ji:1996nm,Radyushkin:1996nd,*Radyushkin:1997ki} .
Deeply virtual meson electroproduction  (DVMP),  specifically for  pseudoscalar meson production, e.g.,  $\eta$ and $\pi^0$, is shown schematically in Fig.~\ref{fig:handbag-pi0}. 

For each quark flavor 
there are  eight leading-twist GPDs. Four correspond to  parton helicity-conserving (chiral-even) processes,  denoted 
by $H^i$,  $\tilde H^i$,  $E^i$ and  $\tilde E^i$, and 
four correspond to parton helicity-flip (chiral-odd) processes  
\cite{Hoodbhoy:1998vm,Diehl:2003ny},  $H^i_T$,  $\tilde H^i_T$,  $E^i_T$ and  $\tilde E^i_T$, 
where 
$i$
denotes quark flavor. 
The GPDs depend on three kinematic variables: $x$, $\xi$ and $t$, where $x$ is the   average longitudinal momentum fraction of  the struck parton before and after the hard interaction and $\xi$ (skewness) is half of the longitudinal momentum fraction transferred to  the struck parton.  Denoting $q$ as the four-momentum transfer and  $Q^2=-q^2$,
the skewness for light mesons of mass $m$, in which  $m^2/Q^2 \ll 1$, can be expressed in terms of the   Bjorken variable $x_B$  as
$\xi\simeq x_B/(2-x_B)$. Here $x_B=Q^2/(2p\cdot q)$ and $t=(p-p^\prime)^2$, where $p$ and $p^\prime$ are the initial and final four-momenta of the nucleon. 
Since the $\pi^0$ and $\eta$  have different combinations of quark flavors, it may be possible 
to  approximately make  a flavor decomposition of the GPDs for up and down quarks.

When the leading order chiral even theoretical calculations for longitudinal virtual photons were compared with the Jefferson Lab $\pi^0$ data \cite{Bedlinskiy:2012be,Bedlinskiy:2014tvi}  
they were  found  to  underestimate the measured cross sections by more than an order of magnitude in their accessible kinematic regions.
The failure to describe the experimental results with quark helicity-conserving operators  stimulated a consideration of the role of the  chiral-odd quark helicity-flip processes. Pseudoscalar meson electroproduction was identified 
as especially sensitive to the quark helicity-flip subprocesses. 
During the past few years, two parallel theoretical approaches -  
\cite{Goloskokov:2009iac,Goloskokov:2011rd}~(GK)  and 
\cite{Ahmad:2008hp} ~(GL) - have been developed utilizing the  chiral-odd GPDs in the calculation of pseudoscalar meson electroproduction. The GL and GK approaches, although employing different models of  GPDs, lead to  transverse photon amplitudes that are much larger than the longitudinal amplitudes. This has been recently confirmed experimentally for  $t$  near $t_{min}$ \cite{Defurne:2016eiy}.

\section{Experimental setup}

The measurements reported here were carried out with the CEBAF Large Acceptance Spectrometer 
(CLAS)~\cite{Mecking:2003zu} 
located in Hall B at Jefferson Lab. The data were obtained   in 2005 in parallel with our previously reported deeply virtual Compton scattering (DVCS) and $\pi^0$ electroproduction experiments 
\cite{Bedlinskiy:2012be,Bedlinskiy:2014tvi,girod:2007jq, Jo:2015ema,DeMasi:2007id}, 
sharing the same physical  setup. 
The integrated luminosity corresponding to the data presented here  
 was $20$ fb$^{-1}$.

\begin{figure}
\includegraphics[width=\columnwidth]{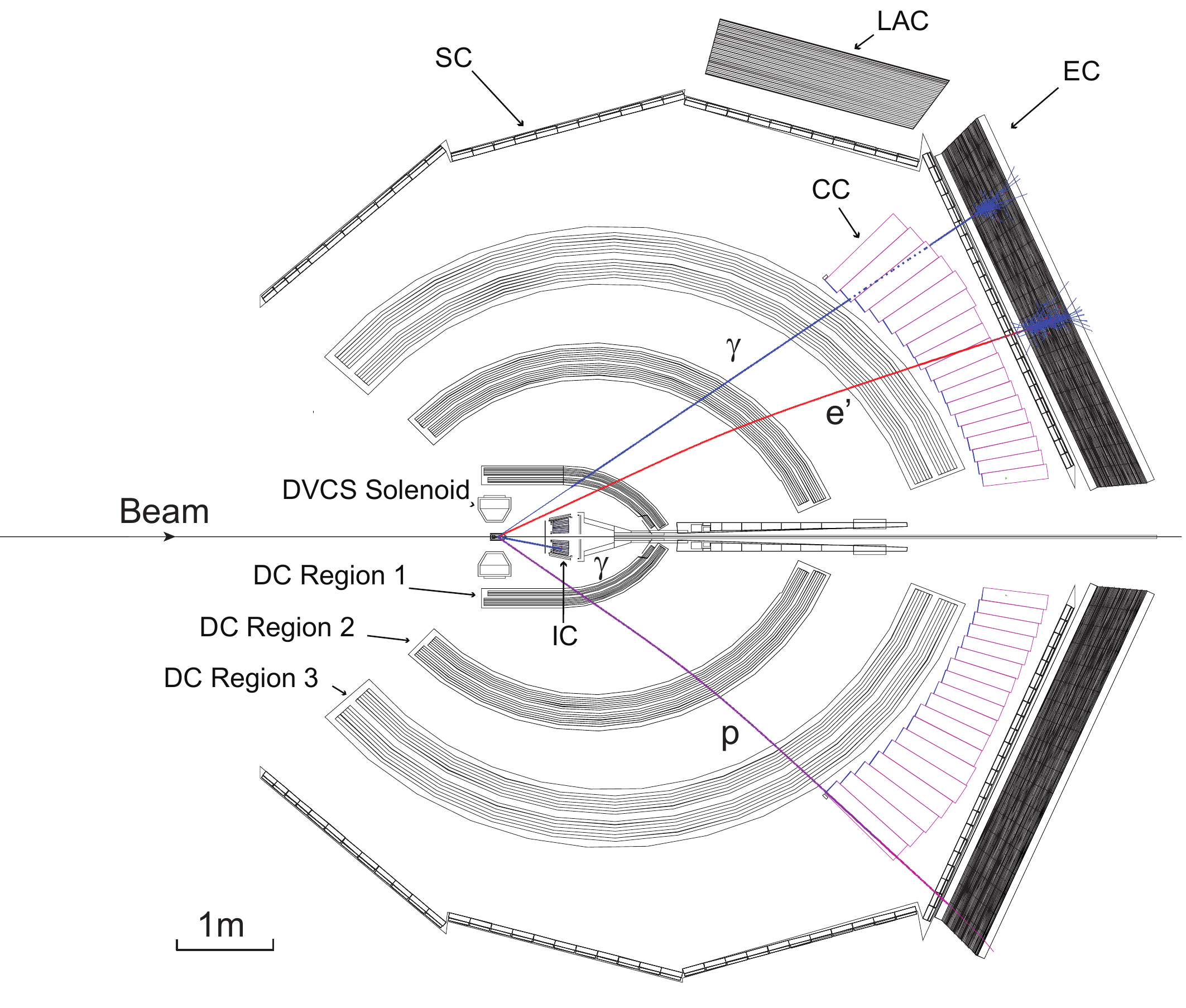}
\caption{(Color online) Schematic view of the CLAS detector in the plane of the beamline constructed by the Monte Carlo simulation program GSIM.  The notation is as follows: inner calorimeter (IC) , electromagnetic calorimeter (EC), large angle electromagnetic calorimeter (LAC), Cherenkov counter (CC), scintillation hodoscope (SC), Drift Chambers (DC).  The LAC was not used in this analysis. The tracks correspond, from top to bottom,  to a photon (blue online), an electron (red online) curving toward the beam line, and a proton (purple online) curving away from the beam line. }
\label{fig:clas}
\end{figure}

The spectrometer consisted of a toroidal-like magnetic field produced by six current coils symmetrically arrayed around the beam axis that divided the detector into six sectors.  The scheme of the CLAS detector array, as coded in the GEANT3-based CLAS simulation code GSIM~\cite{GSIM_User_Guide},
is shown in Fig.~\ref{fig:clas}.

The data were taken using a 5.75 GeV incident electron beam impinging a 2.5 cm long liquid hydrogen target. 
The electron beam was about 80\% polarized. The sign of the beam polarization was changed during measurements at a frequency of 30 Hz. We did not use beam polarization information in this analysis. Effectively, for this experiment the beam was unpolarized.
The target was placed 66 cm upstream of the nominal center of CLAS inside a solenoid magnet to shield the detectors from 
M{\o}ller electrons.

\begin{figure}
\includegraphics[width=\columnwidth]{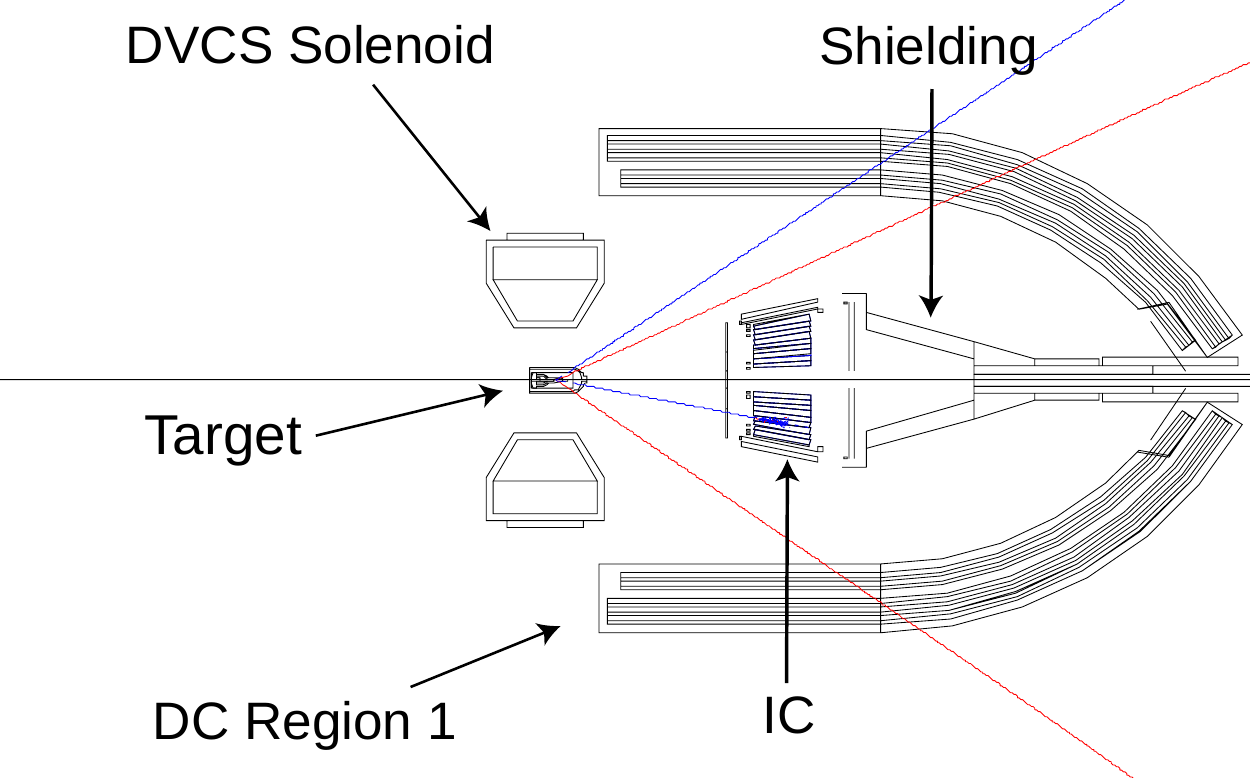}
\caption{(Color online) A blowup of Fig.~\ref{fig:clas} showing  the CLAS target region  in detail. IC is the inner calorimeter and  DC Region~1 represents the drift chambers closest to the target. 
}\label{fig:target}
\end{figure}

Each sector was equipped with three regions of drift chambers (DC)~\cite{Mestayer:2000we}
to determine the trajectory of charged particles, gas threshold Cherenkov counters (CC)~\cite{Adams:2001kk} 
for electron identification, a scintillation hodoscope~\cite{Smith:1999ii}  for time-of-flight (TOF) measurements of charged particles, and an electromagnetic calorimeter (EC)~\cite{Amarian:2001zs}  that was used for electron identification as well as detection of neutral particles. 
To detect photons at small polar angles (from 4.5$^\circ$ up to 15$^\circ$) an inner calorimeter (IC) was added to the standard CLAS configuration, 55 cm downstream from the target. The IC  consisted of 424 PbWO$_4$ tapered crystals whose orientations were projected approximately toward the target.
Figure~\ref{fig:target} zooms in on the target area of Fig.~\ref{fig:clas}
to better illustrate the deployment of the IC and solenoid relative to the target.

The toroidal magnet was operated at a current
corresponding to an integral magnetic field of about 1.36 T-m in the forward direction.
 The magnet polarity was set such that negatively charged particles were bent inward towards the electron beam line. The 
scattered electrons were detected in the CC and EC, which extended from 21$^\circ$ to 45$^\circ$. The lower angle limit was defined by the IC calorimeter, which was located just after the target.

 A Faraday cup was used for the integrated charge measurement with 1\% accuracy. It was  composed of 4000 kg of lead, which corresponds to 75 radiation lengths, and was located 29 m downstream of the target. 

\begin{figure}
\begin{center}
\includegraphics[width=\columnwidth]{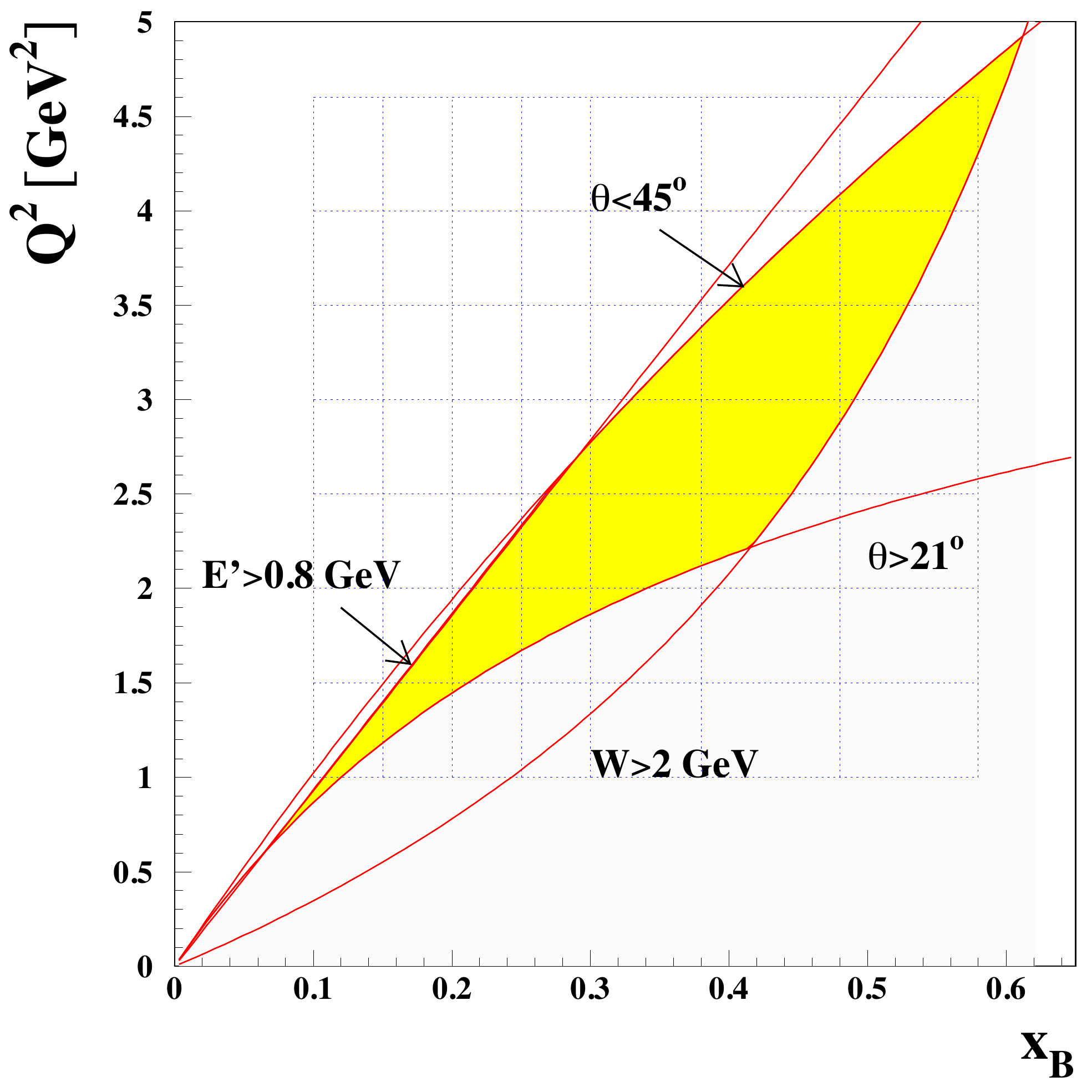}
\caption{(Color online) The kinematic coverage and binning as a function of $Q^2$ and $x_B$. 
The accepted region (yellow online) is determined by the following cuts: 
$W>2$ GeV, $E^\prime>$ 0.8 GeV, $21^\circ<\theta<45^\circ$.
$W$ is the $\gamma^*p$ center-of-mass energy, $E^\prime$ is the scattered electron energy  and $\theta$ is  the  electron's polar angle in the lab frame. The accepted yellow region within each grid boundary represents the kinematic regions for which the cross sections are calculated and presented.} 
\label{fig:kin_cuts}
\end{center}
\end{figure}

\begin{figure*}
\centering
\includegraphics[width=2.25 in]{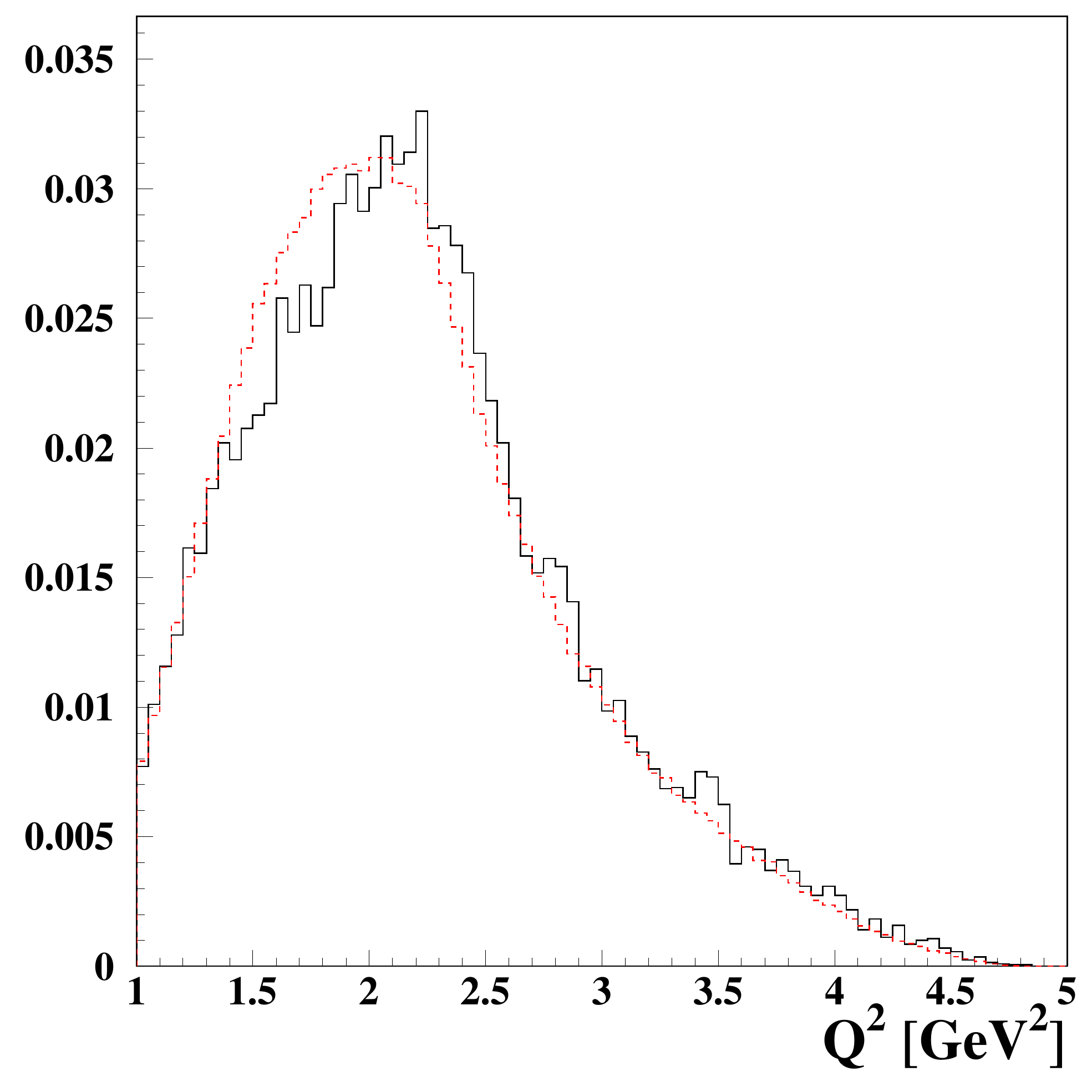}
\includegraphics[width=2.25 in]{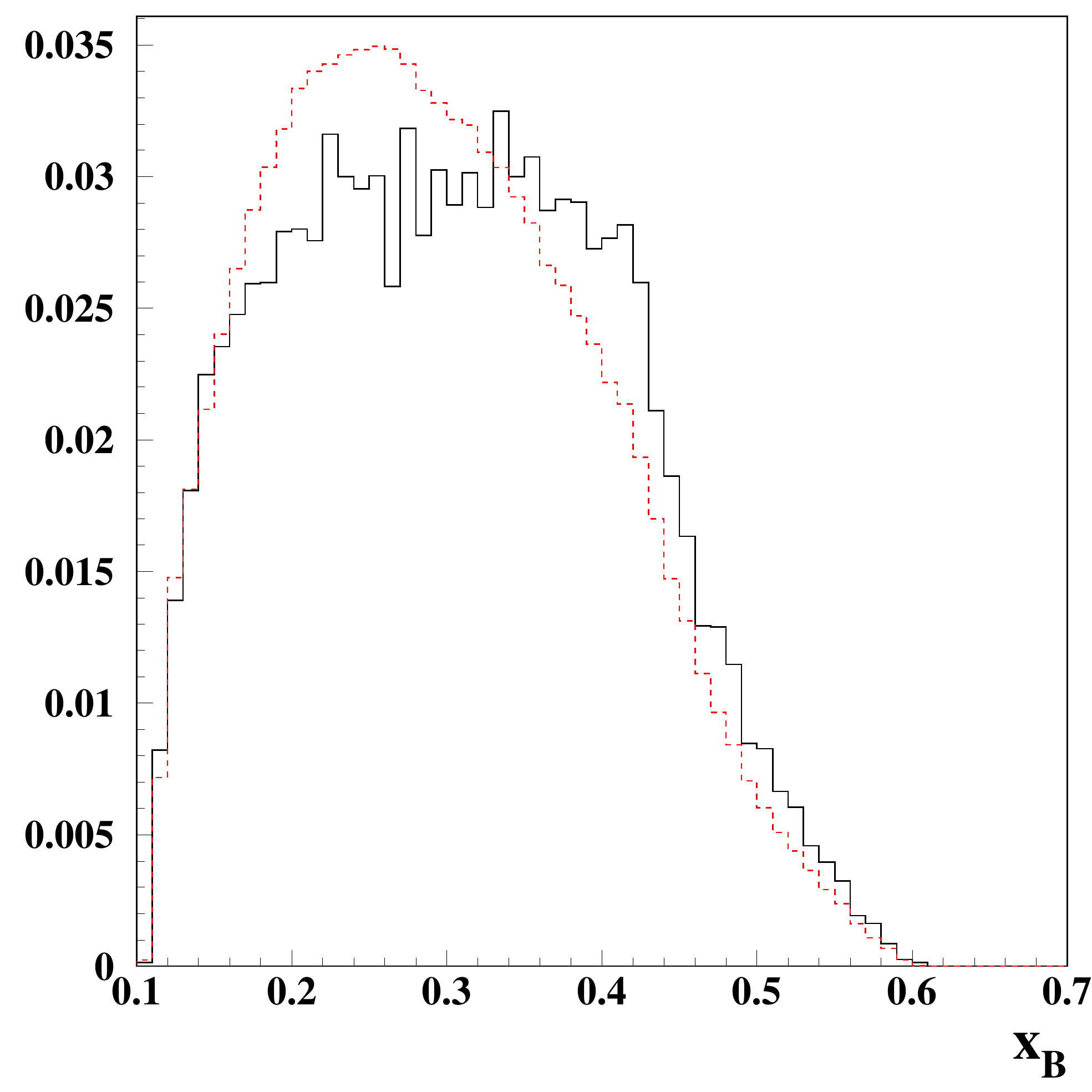}
\includegraphics[width=2.25 in]{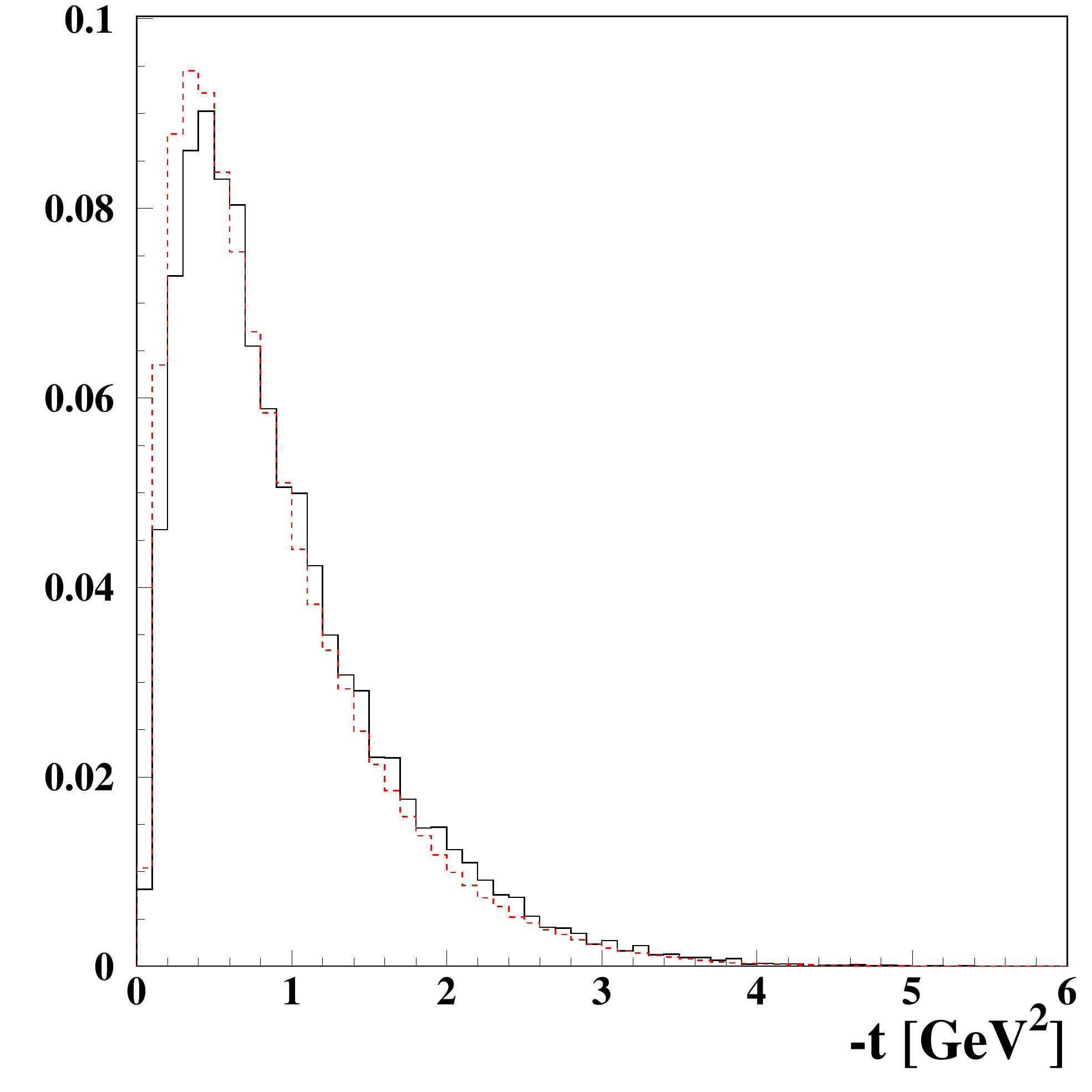}
\includegraphics[width=2.25in]{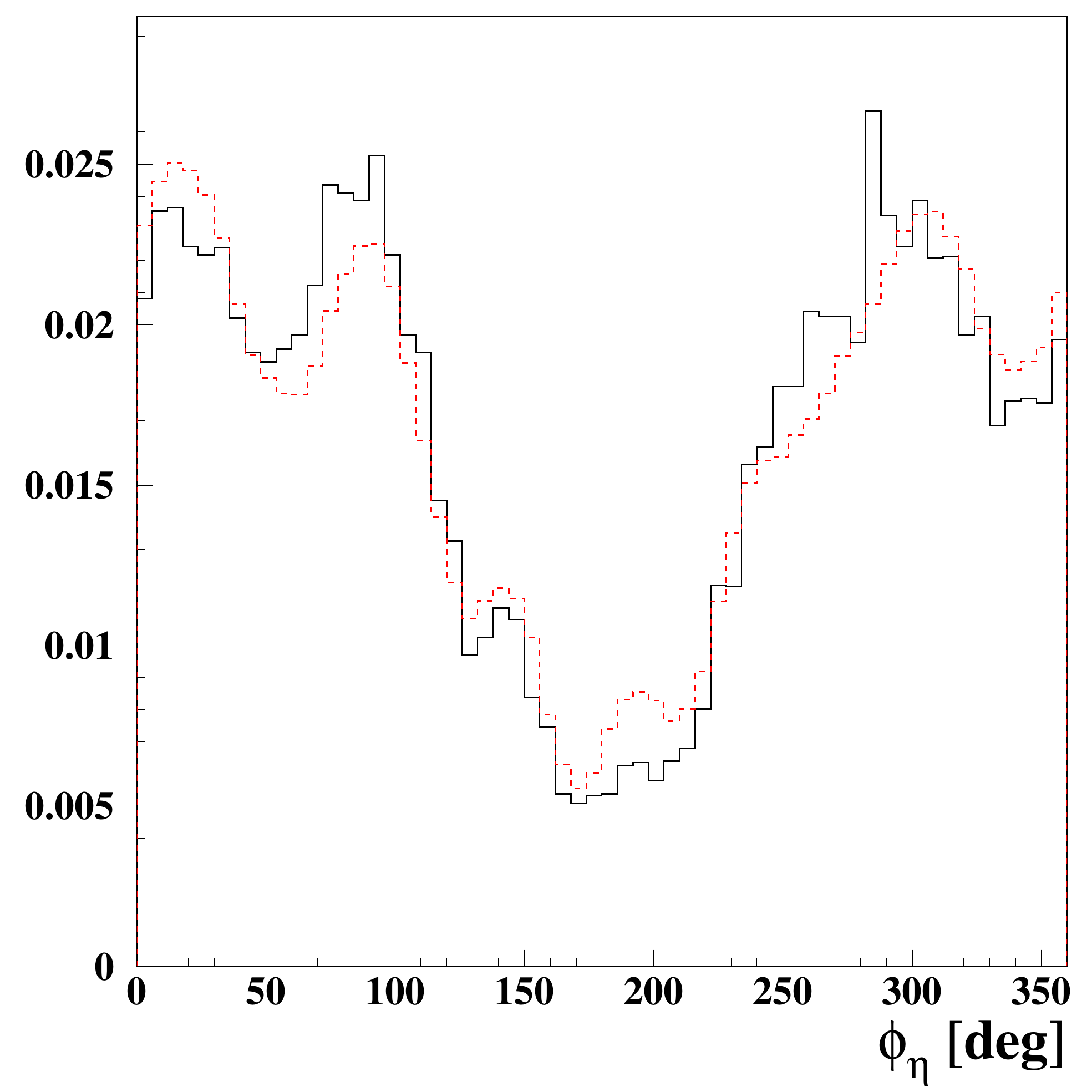}
\caption{(Color online) Yield distributions for kinematic variables $Q^2$, $x_B$, $-t$ and $\phi_\eta$ in arbitrary units. 
The data are in black (solid) and the results of Monte Carlo simulations (see Sec.~\ref{montecarlo}) are in red (dotted). The areas under the curves are normalized to each other. The curves for both the data and Monte Carlo simulations  are the final distributions obtained after tracking and include acceptances and efficiencies. }
\label{fig:kinvar}
\end{figure*}

In the experiment, all four final state particles of the reaction $ep \to e' p' \eta,\  \eta \to \gamma\gamma$ were detected. 
The kinematic coverage for this reaction is shown in Fig.~\ref{fig:kin_cuts},  
and for the individual kinematic variables in Fig.~\ref{fig:kinvar}. For the purpose of physics analysis an additional cut on  $W>2$~GeV was applied as well, where $W$ is the $\gamma^*p$ center-of-mass energy.

The basic configuration of the trigger included the coincidence between signals from   the CC and the EC  in the same sector, with a threshold $\sim 500$ MeV. This was the general trigger for all experiments in this run period.   This threshold  is far  from the kinematic limit of this experiment - $E'>$ 0.8 GeV (see Fig.\ref{fig:kin_cuts}). The accepted region (yellow online) for this experiment is determined by the following cuts: 
$W>2$ GeV, $E^\prime>$ 0.8 GeV, $21^\circ<\theta<45^\circ$. 
Out of a total of about $7 \times 10^9$  recorded  events, about  $20\times10^3$, in 1200 kinematic bins in $Q^2, t, x_B$ and $\phi_\eta$,  for the reaction $ep \to e' p' \eta$, were finally retained. 
The variable $\phi_\eta$ is the azimuthal angle of the emitted $\eta$ relative to the electron scattering plane.

\section{ Particle Identification}

\subsection{Electron identification} 
An electron was identified by requiring the track of a negatively charged particle in the DCs to 
be matched  in space with hits in 
the  CC,  the SC and the  EC.
This electron selection effectively suppresses $\pi^{-}$ contamination up to momenta $\sim$2.5 GeV, which is approximately the threshold for Cherenkov radiation of the $\pi^{-}$ in the CC. 
Additional requirements were used in the offline analysis to refine electron identification and to suppress the  remaining pions.

Energy deposition cuts on the electron signal in the EC  also play an important role in suppressing the pion background.
 An electron propagating through the  calorimeter produces an electromagnetic shower and deposits a large fraction of its energy in the calorimeter proportional to its momentum, while  pions typically lose a smaller fraction of their energy,  primarily   by ionization.

\begin{figure}[h]
\centering 
\includegraphics[width=0.4\textwidth]{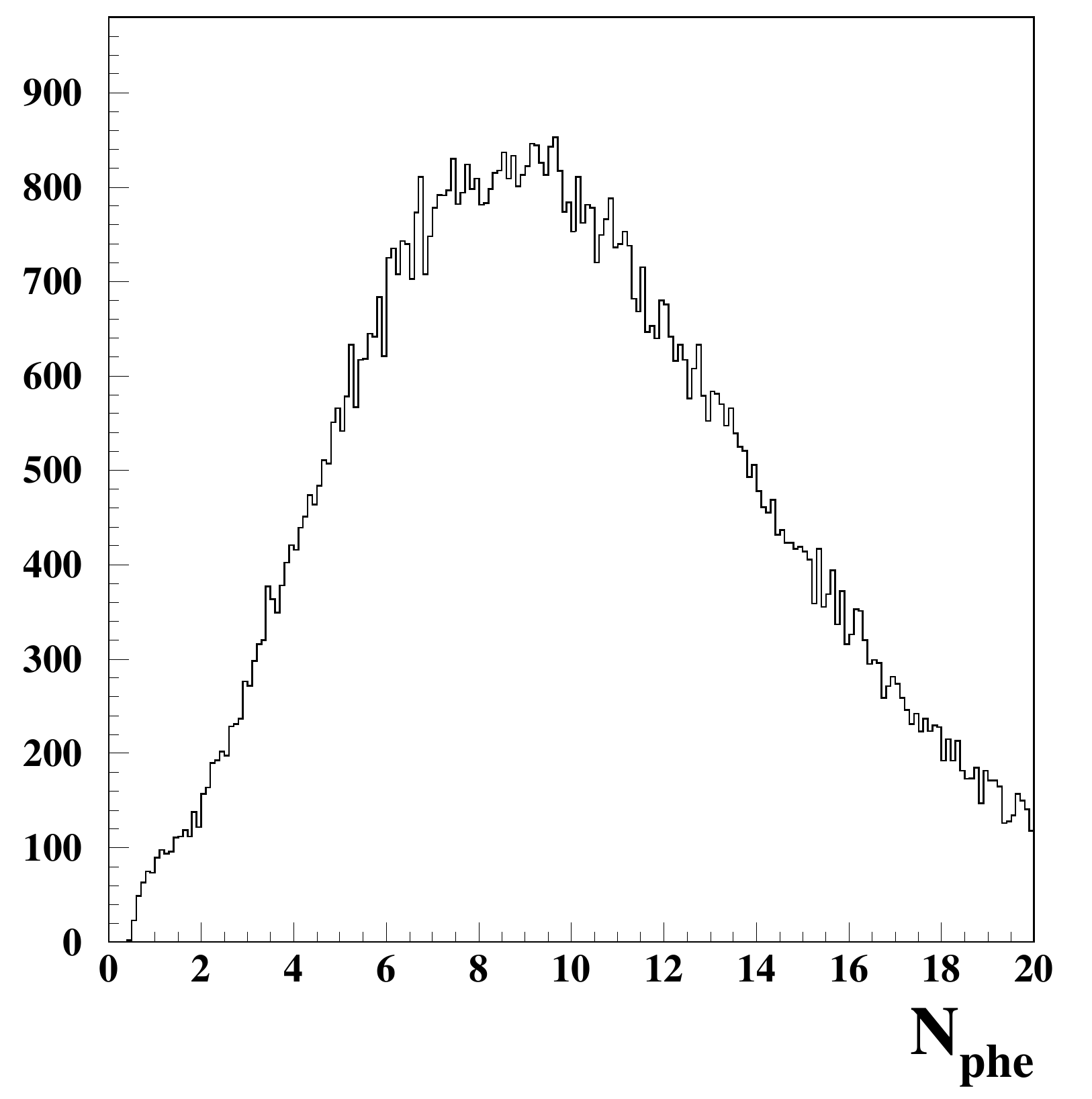}
\caption{
The number of CC photoelectrons for events that pass all cuts.}
\label{fig:cc_match}
\end{figure}

The distribution of the number of the photoelectrons in the CC after all selection criteria were applied is shown in 
Fig.~\ref{fig:cc_match}. 
The residual small shoulder around
$N_{phe}=1$ represents the  pion contamination, which is seen to be negligibly small after applying  all selection criteria. 

\begin{figure*}
\subfigure[Before any exclusivity cuts.]
{\includegraphics[width=\columnwidth]{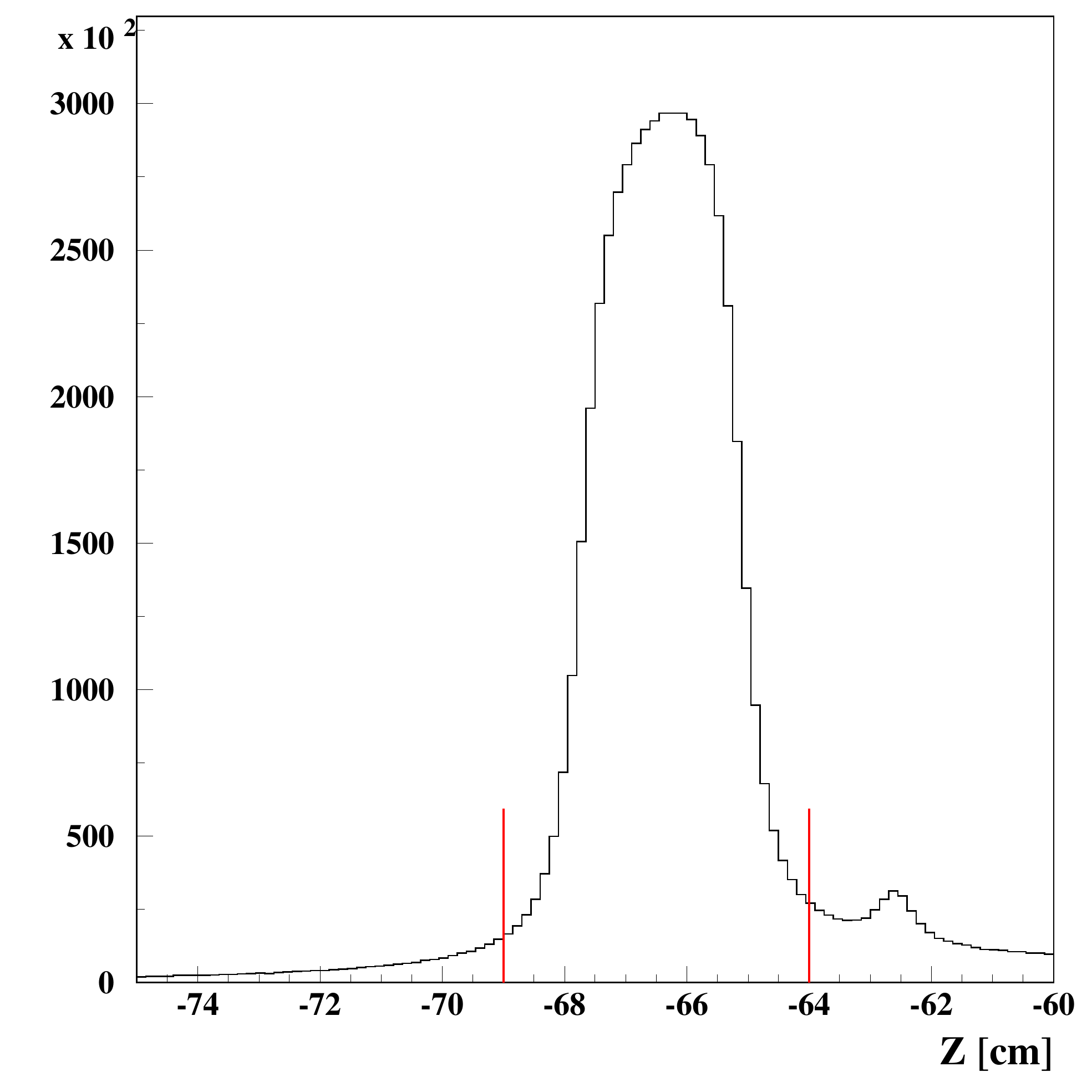}}
\subfigure[After $\eta$ exclusivity cuts.]
{\includegraphics[width=\columnwidth]{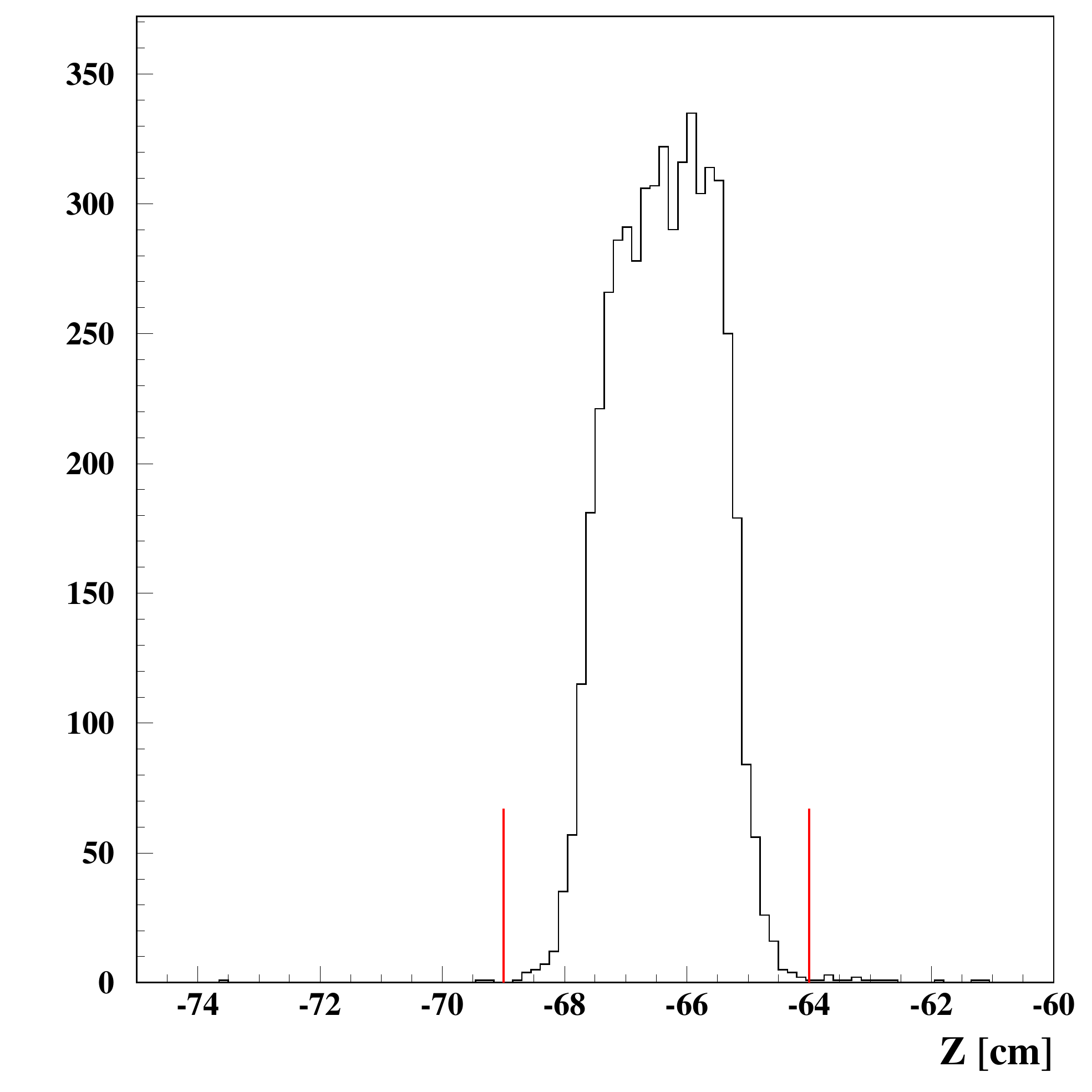}}
\caption{
The $z$-coordinate of the electron vertex. The vertical lines are the positions of the applied cuts. Note in (a) the small peak to the right of the target that is due to a foil placed at  $z=-62.5$ cm downstream of the target window.
In panel (b) the peak due to the foil  disappears after the selection of the exclusive reaction.}
\label{fig:vertex2}
\end{figure*}

The charged particle tracks were reconstructed by the drift chambers. The vertex location was calculated from the intersection of the track with the beam line. 
 A cut was applied on the  $z$-component of the electron vertex  position to eliminate events  originating outside the target. The vertex distribution and cuts for one of the sectors are shown in Fig.~\ref{fig:vertex2}. The left plot shows the $z$-coordinate distribution before the exclusivity cuts, which are described below in Section~\ref{sect:exclusivity_cuts},  and the right plot  is the distribution after the exclusivity cuts. The peak at  $z=-62.5$ cm exhibits the interaction of the beam with an insulating foil,
which is completely removed after the application of the exclusivity cuts, demonstrating that these cuts very effectively exclude  the interactions involving nuclei of the surrounding nontarget material.

\subsection{Proton identification} 
The proton was identified as a positively charged particle with the correct time-of-flight.
The  quantity of interest  ($\delta t=t_{SC}-t_{exp}$) is the difference in the time between the measured flight time from the event vertex to the SC system ($t_{SC}$) and that expected for the proton ($t_{exp}$). The quantity $t_{exp}$ was computed from the velocity of the particle and the track length. The velocity was determined from the momentum by assuming the mass of the  particle equals that of a proton.
 A cut at the level of $\pm 5 \sigma_t$ was applied around  $\delta t = 0$, where  $\sigma_t$  is the time-of-flight resolution, which is momentum dependent.
This wide  cut was  possible because the exclusivity cuts (see Section IV B below) very effectively suppressed the remaining pion contamination. 

\subsection{Photon identification} 

Photons were detected in both calorimeters, the EC  and IC. 
In the EC, photons  were identified as {\it neutral} particles  with $\beta>0.8$ and $E>0.35$ GeV. 
Fiducial cuts were applied to avoid  the EC  edges. When a photon hits the boundary of the calorimeter, the energy cannot be fully reconstructed due to the leakage of the shower out of the detector.
Additional fiducial cuts on the EC  were applied to account for the shadow of the IC  (see Fig.~\ref{fig:clas}). 
The  calibration of the EC  was done using cosmic muons and  the photons from neutral pion decay ($\pi^0\to\gamma\gamma$).

In the IC,   each detected cluster was considered  a photon. The assumption was made that this photon originated from the electron vertex. Additional geometric cuts were applied to remove low-energy clusters around the beam axis and photons near the edges of the IC, where the energies of the photons were incorrectly reconstructed due to the electromagnetic shower leakage.
The photons from $\eta\to \gamma\gamma$ decays were detected in the IC  in an angular range between $5^\circ$  and $17^\circ$   and in the EC   for angles greater than $21^\circ$. 
The reconstructed invariant mass of two-photon events was then subjected to various cuts to isolate exclusive $\eta$ events, with a  residual background, as discussed in 
Section IV B below.  

\subsection{Kinematic corrections}
Ionization energy-loss corrections were applied to  protons and electrons in both data and Monte Carlo events. These corrections were estimated using the GSIM Monte Carlo program.
Due to imperfect knowledge of the properties of the CLAS detector, such as the magnetic field distribution and the precise  placement of the components or detector materials, small empirical sector-dependent corrections had to be made on the momenta and angles of the detected electrons and protons. The corrections were determined by systematically studying the kinematics of the particles emitted from well understood kinematically-complete processes, e.g., elastic electron scattering. These corrections were on the order of 1\%.

\section{Event selection}

\subsection{Fiducial cuts}
Certain areas of the detector acceptance were not efficient due to  gaps in the DC,
problematic SC counters, and inefficient zones of the CC and the EC. These areas were removed from the analysis as well as from the simulation by means of geometrical cuts,  which were momentum, polar angle and azimuthal angle dependent.

In addition, we excluded events,  when a photon from the $\eta$-decay or
Bremsstrahlungs photon was detected in the same sector as the electron. This avoids additional photons  which are close in space to the scattered lepton leaving
a signal in the EC close to where the supposed lepton hits the EC. This was done for both the experimental data as well as the Monte Carlo data used for
correcting experimental yields.

\subsection{Exclusivity cuts}
\label{sect:exclusivity_cuts}

To select the exclusive reaction $ep\rightarrow e'p'\eta$, each event was required to contain an electron, one proton and at least two photons in the final state. Then,  so called {\it exclusivity cuts} were applied to all combinations of an electron,  a proton and two  photons  to ensure energy and momentum conservation, thus eliminating events in which there were any additional undetected particles.

\begin{figure}
\includegraphics[width=\columnwidth]{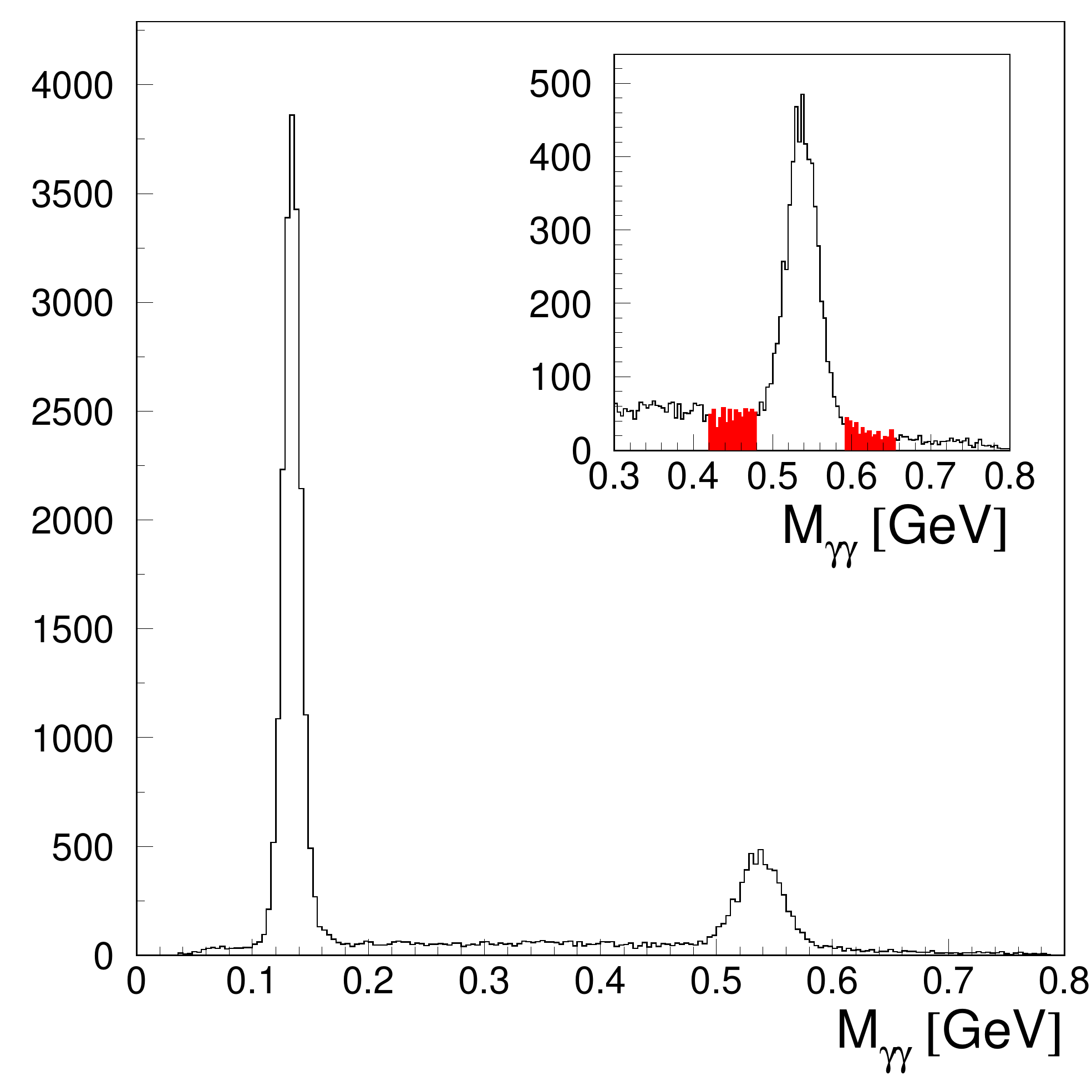}
\caption{
\label{Mgg}
(Color online)  The two-photon invariant  mass distribution, $M_{\gamma\gamma}$, after all exclusivity cuts have been applied, 
for the case where the two photons are detected by the IC. The large peak at lower $M_{\gamma\gamma}$ is due to $\pi^0$ electroproduction and the smaller peak at higher 
$M_{\gamma\gamma}$ is due to $\eta$ electroproduction. 
The inset magnifies the region around the $\eta$ peak. The filled regions above and below the peak (red online) are the sidebands that are used for background subtraction, as discussed in the text.
}
\end{figure}

Four cuts were used for the exclusive event selection \begin{itemize}

\item   $\theta_X<2^o$, where   $\theta_X$ is  the angle between the reconstructed  
$\eta$ momentum vector and the missing momentum vector for the reaction $ep\to e'p'X$.
\item the missing mass squared $M_x^2(e'p')$ of the $e'p'$ system  ($ep\to e'p'X$), with $| M_x^2(e'p') - M^2_{\eta}| < 3\sigma$;
\item the missing mass $M_x(e'\gamma\gamma)$ of the $e'\gamma\gamma$ system  ($ep\to e'\gamma\gamma X$), with $|M_x(e'\gamma\gamma)-M_{p}| < 3\sigma$;
\item the missing energy  $E_x(e'p'\eta)$ ($ep\to e'p'\gamma\gamma X$), with $|E_x(e'p'\eta)- 0| < 3\sigma$;
\end{itemize}
\noindent
Here $\sigma $ is the observed experimental resolution obtained as the standard deviation from the mean value of the distributions of each quantity. Three sets of resolutions were determined independently for each of the three photon-detection topologies (IC-IC, IC-EC, EC-EC). The invariant mass $M_{\gamma\gamma}$ for the two detected photons, where both photons were detected in the IC, after these cuts  is shown in Fig.~\ref{Mgg}. The two peaks correspond to  $\pi^0$ and $\eta$ production, with the $\pi^0$ production exhibiting a significantly larger  cross section than  $\eta$ production. The distributions were generally broader than  in the Monte Carlo simulations so that the cuts for the data were typically broader than those used for the Monte Carlo simulations.
Similar results were obtained for the topology in which one photon was detected in the IC and one in the EC, as well as the case where both photons were detected in the EC.

\subsection{Background subtraction}

The $M_{\gamma\gamma}$ distribution contains  background under the $\eta$ peak even after the application of all exclusivity cuts shown in the insert of Fig.~\ref{Mgg}.
The background under the $\eta$ invariant mass peak was subtracted for each kinematic bin.
It was found that most of the  background  comes from  
the production of  $\pi^0$ meson, together  with the detection of only one decay photon 
 with an accidental photon signal in the electromagnetic calorimeter.
Thus, the background was subtracted using  the following procedure. All $\pi^0$ events which were in coincidence with  accidental photons were identified.  Then, the distributions of the invariant masses of one of the  $\pi^0$ decay photons with the accidentals  were obtained, and normalized with respect to the side bands around the $\eta$ mass. The sidebands  were determined as $(-6\sigma,-3\sigma) \cup  (3\sigma,6\sigma)$ in the $M_{\gamma\gamma}$ distributions, as shown in Fig.~\ref{Mgg}.
 
The   resulting events in the region between side bands were then subtracted as the background contamination.   The mean ratio of   background to peak over all kinematic bins and all  combinations of IC and EC is about 25\%.

 \subsection{Kinematic binning}
 
 The kinematics of the reaction  are defined by four variables: $Q^2$, $x_B$,  $t$ and $\phi_{\eta}$.
In order to obtain differential cross sections the data were divided into four-dimensional rectangular bins in these variables.
There are seven bins in $x_B$, seven bins in $Q^2$ as shown 
in Tables \ref{q2-bins}--\ref{x-bins} and 
in Fig.~\ref{fig:kin_cuts}. For each $Q^2$-$x_B$ bin there are nominally eight bins in $t$
(Table ~\ref{t-bins}),
but the actual number  is determined by the  kinematic acceptance in $t$ 
for each $Q^2$-$x_B$ bin, as well as the available statistics. Differential cross section distributions were obtained for 20 bins in $\phi_\eta$ for each kinematic bin in  $Q^2$, $x_B$ and $t$.

\begingroup
\squeezetable
\begin{table}
\caption{$Q^2$ bins}
\begin{ruledtabular}
\begin{tabular}{ccc}
Bin Number & Lower Limit & Upper limit \\ 
 & (GeV$^2$)   & (GeV$^2$) \\ \hline
1 & 1.0   &  1.5 \\ 
2 & 1.5   &  2.0 \\   
3 & 2.0   &  2.5 \\
4 & 2.5   &  3.0 \\ 
5 & 3.0   &  3.5 \\ 
6 & 3.5   &  4.0 \\  
7 & 4.0   &  4.6 \\
\end{tabular}
\label{q2-bins}
\end{ruledtabular}

\caption{$x_B$ bins}
\begin{ruledtabular}
\begin{tabular}{ccc}
Bin Number & Lower Limit & Upper limit \\ \hline
1 & 0.10   &  0.15 \\
2 & 0.15   &  0.20 \\  
3 & 0.20   &  0.25 \\
4 & 0.25   &  0.30 \\ 
5 & 0.30   &  0.38 \\
6 & 0.38   &  0.48 \\  
7 & 0.48   &  0.58 \\ 
\end{tabular}
\end{ruledtabular}
\label{x-bins}

\caption{$|t|$ bins}
\begin{ruledtabular}
\begin{tabular}{ccc}
Bin Number & Lower Limit & Upper limit \\ 
           & (GeV$^2$)   & (GeV$^2$) \\ \hline

1 & 0.09   &  0.15 \\ 
2 & 0.15   &  0.20 \\   
3 & 0.20   &  0.30 \\ 
4 & 0.30   &  0.40 \\
5 & 0.40   &  0.60 \\
6 & 0.60   &  1.00 \\ 
7 & 1.00   &  1.50 \\
8 & 1.50   &  2.00 \\
\end{tabular}
\end{ruledtabular}
\label{t-bins}
\end{table}
\endgroup

\section{Cross sections for $\gamma^*p\to \eta p^\prime$}
The fourfold differential cross section as a function of the four variables $(Q^2,x_B,t ,\phi_\eta)$ was obtained from the expression

\begin{equation}
\begin{split}
\frac{d^4 \sigma_{ep \rightarrow e^\prime p^\prime \eta}}{dQ^2 dx_B dt d\phi_\eta} = 
 \frac{N(Q^2,x_B,t,\phi_\eta)}{ \Delta Q^2  \Delta x_B \Delta t \Delta \phi_\eta} \times \\
\frac{1}{\mathcal{L}_{int}\epsilon_{ACC}\delta_{RC} \delta_{Norm}Br(\eta\to\gamma\gamma)}.
\end{split}
\label{eq:sig_ep_eppippim}
\end{equation}

\noindent The definitions of the  quantities in Eq.~\ref{eq:sig_ep_eppippim}  are:

\begin{itemize}

\item $N(Q^2,x_B,t,\phi_\eta)$ is the  number of $ep \rightarrow e^\prime p^\prime \eta$ events in a given ($Q^2,x_B,t,\phi_\eta$) bin;

\item $\Delta Q^2 \Delta x_B \Delta t \Delta \phi_\eta$ is the corresponding 4-dimensional bin volume. 
The accepted kinematic bin volumes in   $Q^2,   x_B,  t,  {\text{ and }}  \phi_\eta$ are typically smaller than 
the product 
$\Delta Q^2 \cdot \Delta x_B \cdot \Delta t \cdot \Delta\phi_\eta$ of the 4-dimensional grid 
because of cuts in $\theta_e$, $W$ \text {and} $E^\prime$ (e.g. see Fig.~\ref{fig:kin_cuts} ). 
The reported $Q^2$, $x_B$ \text {and} $t$  value for each bin is the mean value of the accepted volume assuming a constant density of events.

\item $\mathcal{L}_{int}$ is the integrated luminosity (which takes
into account the correction for the data-acquisition dead time);
 
\item $\epsilon_{ACC}$ is the acceptance calculated for each bin $(Q^2,x_B,t,\phi_\eta)$
(see Sec.~\ref{montecarlo})
;

\item $\delta_{RC}$ is the correction factor due to the radiative effects calculated for each $(Q^2,x_B,t,\phi_\eta)$ bin
(see Sec.~\ref{rccorrection})
;

\item $\delta_{Norm}$ is the overall absolute normalization factor calculated from the elastic cross section measured in the same experiment 
(see  Sec.~\ref{normalization});
 
\item $Br(\eta\to\gamma\gamma)=\frac  {\Gamma(\eta\to\gamma\gamma) }   {\Gamma_{total}}=0.394$ 
~\cite{Agashe:2014kda} 
is the branching ratio for the $\eta \to \gamma\gamma$ decay mode.

\end{itemize}

The reduced or ``virtual photon"  cross sections  were extracted from the fourfold cross section (Eq. \ref{eq:sig_ep_eppippim}) through:

\begin{equation}
\frac{d^2\sigma_{\gamma^* p \rightarrow p^\prime \eta} }{dt d\phi_\eta} = 
\frac{1}{\Gamma_V(Q^2,x_B,E)} 
\frac{d^4\sigma_{ep\rightarrow e^\prime p^\prime \eta}}{dQ^2 dx_B dt d\phi_\eta}.
\end{equation}

\noindent
The Hand convention~\cite{Hand:1963bb} was adopted for the
definition of the virtual photon flux $\Gamma_V$: 

\begin{equation}
\Gamma_V (Q^2,x_B,E)= \frac{\alpha}{8\pi} \frac{Q^2}{m_p^2 E^2} 
\frac{1-x_B}{x_B^3} \frac{1}{1-\epsilon},
\label{eq:GammaV}
\end{equation}
\noindent where $\alpha$ is the standard electromagnetic coupling constant.
The variable $\epsilon$ represents the ratio of fluxes of longitudinally  and transversely polarized virtual photons and is given by
\begin{equation}
\epsilon=\frac {1-y-\frac{Q^2}{4E^2}}   {1-y+\frac{y^2}{2}+\frac{Q^2}{4E^2} }
\label{epsilon},
\end{equation}
with $y=p \cdot q/q \cdot k=\nu/E$.

\begin{figure}
\includegraphics[width=\columnwidth]{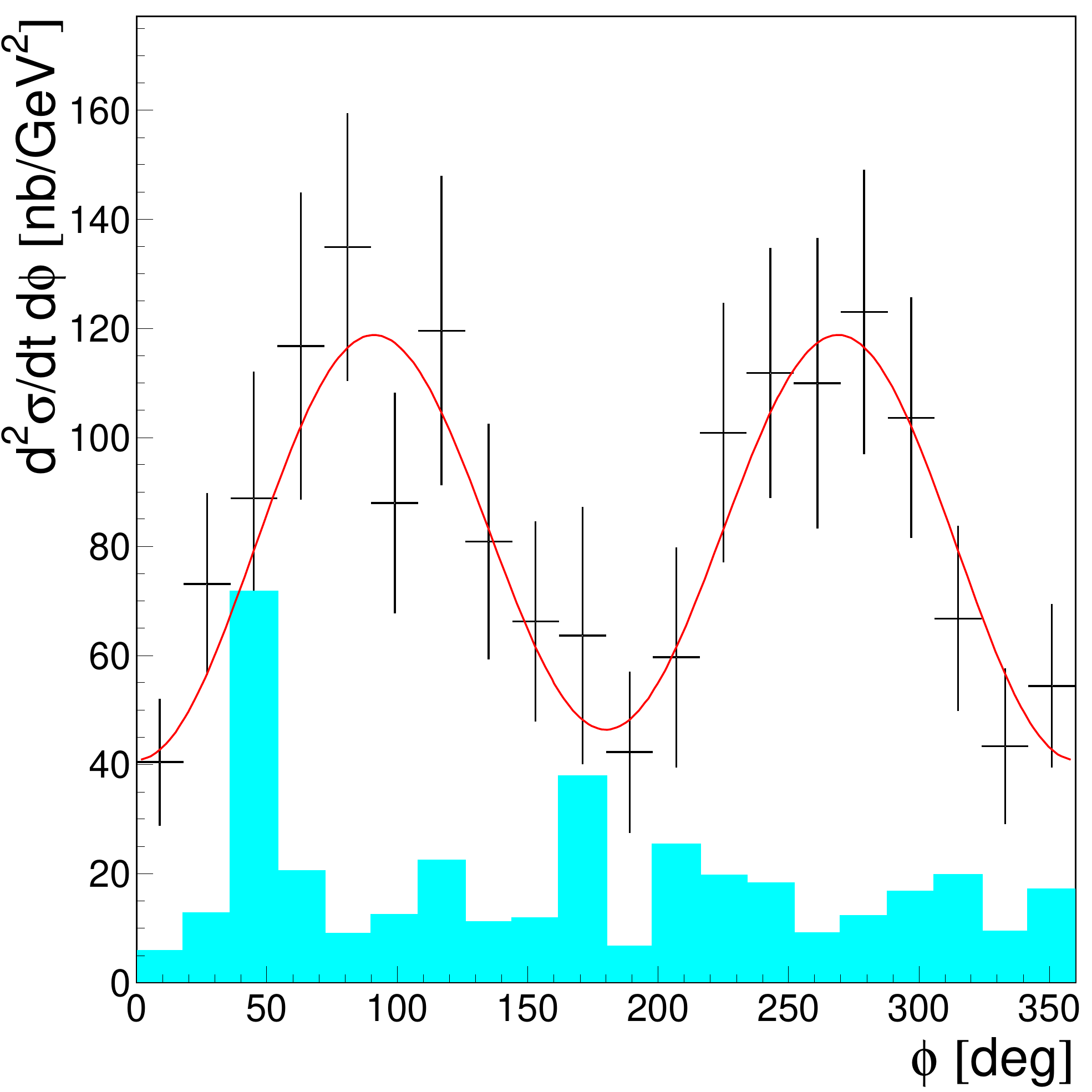}
\caption{
(Color online)
The differential cross section $d^2\sigma/dtd\phi_\eta$ for the reaction $\gamma^*p\to p^\prime\eta$ for the kinematic interval at 
$Q^2=1.75$ GeV$^2$, $x_B=0.23$ and $t=-0.8$ GeV$^2$. 
The error bars indicate statistical uncertainties. 
Systematic uncertainties are indicated by the cyan bars.
The red curve is a fit 
in terms of the structure functions in Eq.~\ref{eq:sigmaphidep}. 
} 
\label{fig:cross-section} 
\end{figure}

A table of  the  reduced cross sections  can be obtained online in Ref. \cite{Supplemental_eta-PRC}.
An example of the differential cross section as a function of 
$\phi_\eta$ in a single kinematic interval in $Q^2, t$ and $x_B$ is shown in Fig.~\ref{fig:cross-section}.

\section{Monte Carlo simulation}
\label{montecarlo}
The acceptance for each ($Q^2$, $x_B$, $t$, $\phi_\eta$) bin of the CLAS detector with the present setup for the reaction $ep\rightarrow e'p'\gamma\gamma$ was calculated using the Monte Carlo program GSIM. The event generator used an empirical parametrization of the cross section as a function  of $Q^2$, $x_B$ and $t$. The parameters were tuned using the MINUIT program to best match the  simulated $\eta$ cross section with the measured  electroproduction cross section. 
Two iterations were found to be sufficient to describe the experimental cross section and distributions.  The comparisons of the experimental and Monte Carlo simulated distributions are shown in Fig.~\ref{fig:kinvar} for the variables $Q^2$, $x_B$, $-t$ and $\phi_\eta$.

Additional smearing factors for tracking and timing resolutions were included in the simulations to provide more realistic resolutions for charged particles. The Monte Carlo events were analyzed by the same code that was used to analyze the experimental data, and with the additional smearing and somewhat different exclusivity cuts, to account for the leftover discrepancies in calorimeter resolutions. Ultimately the number of reconstructed Monte Carlo events  was an order of magnitude higher  than the number of reconstructed experimental events.  Thus, the statistical uncertainty introduced by the acceptance calculation was typically much smaller than the statistical uncertainty of the data.

The efficiency of the event reconstruction depends on the level of noise in the detector, the greater the noise the lower the efficiency. It was found that the efficiency for reconstructing particles decreased linearly with increasing beam current.
To take this into account the background hits from random 3-Hz-trigger events  were mixed with the Monte Carlo events for all detectors - DC, EC, IC, SC and CC.
The acceptance for a given bin  was calculated as a ratio of the number of reconstructed events to the number of generated events  as

\begin{equation}
\epsilon_{ACC}(Q^2,x_B,t,\phi_\eta)=\frac {N^{rec}(Q^2,x_B,t,\phi_\eta)}{N^{gen}(Q^2,x_B,t,\phi_\eta)}.
\end{equation}

\noindent
 Only areas of the 4-dimensional space with an acceptance equal to or greater than 0.5\% were used.
This cut was applied to avoid the regions where the calculation of the acceptance was not reliable. 
\section{Radiative Corrections}
\label{rccorrection}

\begin{figure}
\includegraphics[width=2.cm]{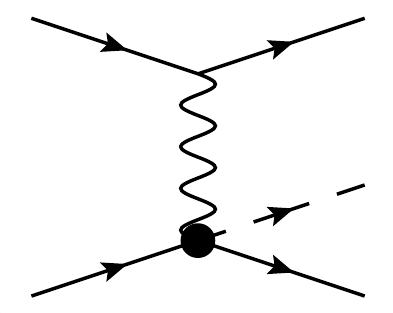}
\includegraphics[width=2.cm]{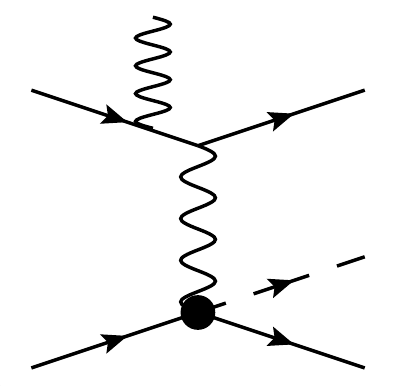}
\includegraphics[width=2.cm]{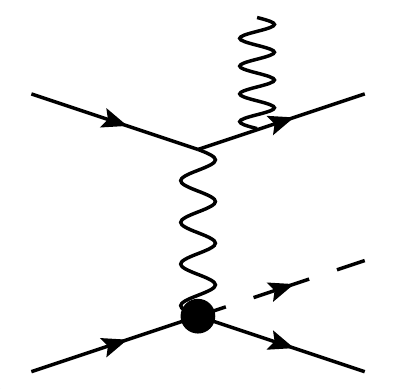}
\includegraphics[width=2.cm]{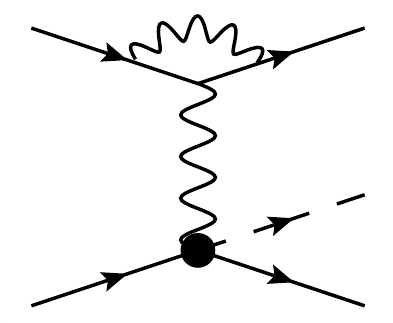}
\includegraphics[width=2.cm]{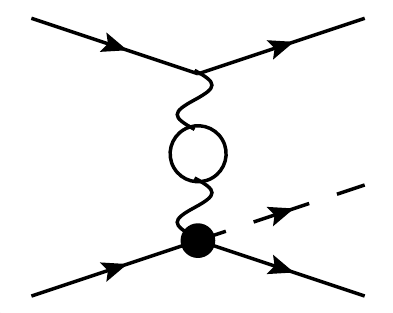}
\caption{\label{fig:rad_proc}	
Feynman diagrams contributing to the $\eta$ electroproduction
cross section. Left to right: Born process,
Bremsstrahlung (by the initial and the final electron),  vertex correction, and
vacuum polarization.}
\end{figure}

The QED processes include radiation of  photons  that are not detected by  the experimental set up, as well as vacuum polarization 
and lepton-photon vertex corrections (see Fig.~\ref{fig:rad_proc}).
These processes can be calculated  from QED and the measured
cross section can be corrected for these effects \cite{Mo:1968cg}. The radiative corrections, $\delta_{RC}$, for the experiment are give by
\begin{equation}
	\sigma_{\eta} = \frac{\sigma_\eta^{meas}}{\delta_{RC}}.
\end{equation}

\noindent
Here $\sigma_\eta^{meas}$ is the observed cross section  and 
$\sigma_{\eta} $ is the $\eta$ electroproduction cross section after corrections.

The radiative corrections were obtained using the software package EXCLURAD \cite{Afanasev:2002ee}, which has been  used for radiative corrections in previous CLAS experiments. 
The same analytical structure functions were implemented in the EXCLURAD package as were used to generate the $\eta$ electroproduction events in the Monte Carlo simulation. The corrections were  computed for each kinematic bin of  $Q^2$, $x_B$,   $t$ and $\phi_\eta$. 
\begin{figure}
\includegraphics[width=\columnwidth]{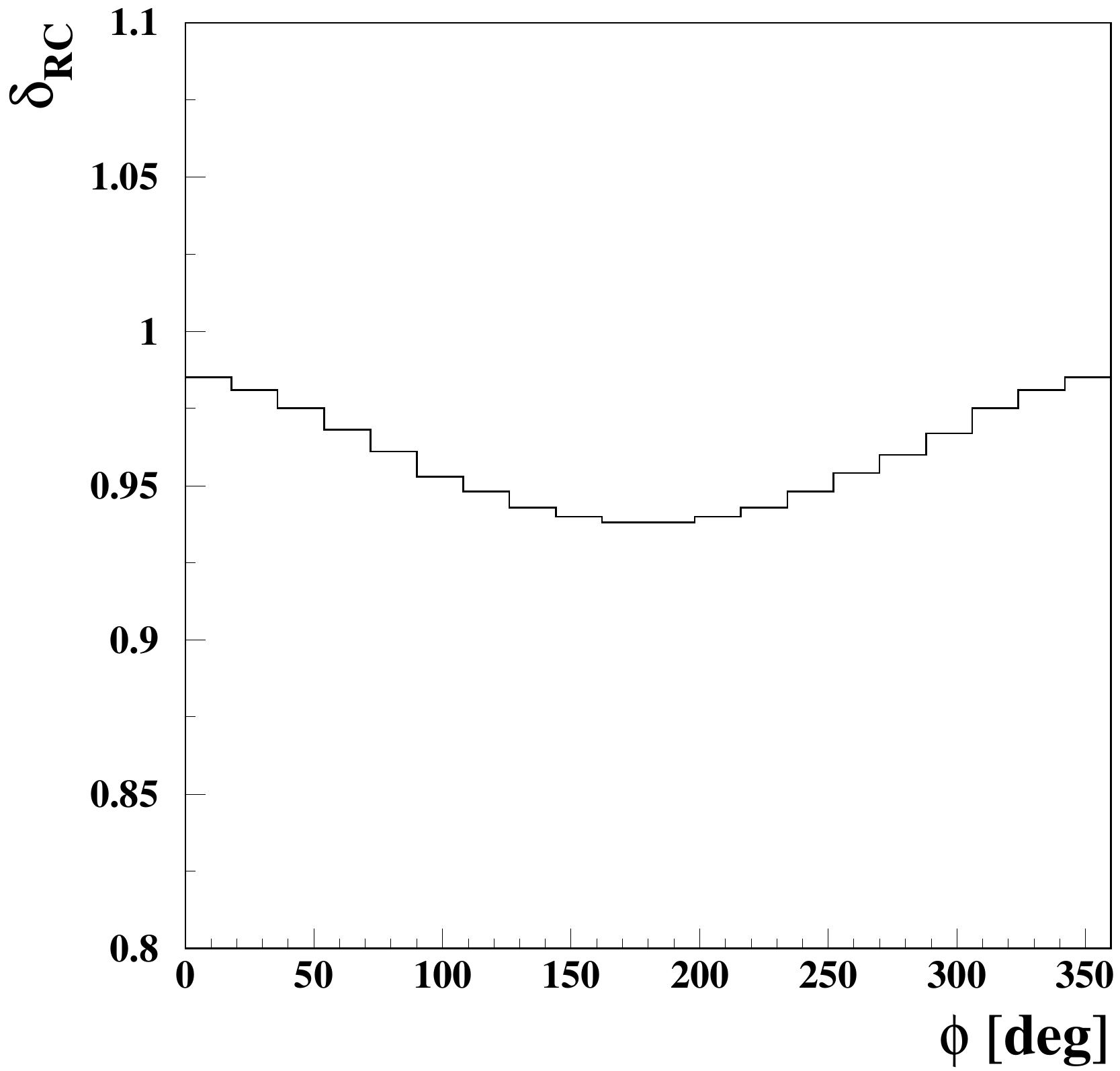}
\caption{Radiative corrections  $\delta_{RC}$  for $\eta$  electroproduction as a function of $\phi_\eta$ 
for the kinematic interval at 
$Q^2=1.15$ GeV$^2$, $x_B=0.13$ and $t=-0.12$ GeV$^2$. 
}
\label{fig:rc_2d}
\end{figure}
Figure ~\ref{fig:rc_2d} shows the radiative corrections   
for the first kinematic bin $(Q^2,x_B,t)$ as a function of the $\phi_\eta$.

\section{Normalization Correction}
\label{normalization}
To check the overall absolute normalization,   the cross section of  elastic electron-proton scattering was measured using the same data set.  The measured cross section was lower than the known elastic cross section \cite{Bosted:1994tm,Christy:2004rc} by approximately 13\% over most of the elastic kinematic range. Studies made using additional other reactions where the cross sections are well known, such as $\pi^0$ production in the resonance region, and Monte Carlo simulations of the effects of random backgrounds, indicate that the measured cross sections were $\sim$13\% lower
than the available published cross sections over a wide kinematic range.
Thus, a normalization factor $\delta_{Norm}\sim 0.87$ was  applied to the measured cross section. This value includes the efficiency of the SC counters, which was estimated to be around  95\%, as well as other efficiency factors that are not accounted for in the analysis, such as trigger and CC efficiency effects.

\section{Systematic Uncertainties}
\label{systematics}
There are various sources of systematic uncertainties. Some  are introduced in the analysis, while others can be tracked back to uncertainties of measurements such as target length or integrated luminosity.  Still others are related to an imperfect knowledge of the response of the spectrometer. In most cases uncertainties originating from the analysis itself can be estimated separately for each kinematic bin ($Q^2$,$x_B$,$t$,$\phi_\eta$). Where bin-by-bin estimates are not possible, global values for all bins are estimated. 

A source of systematic uncertainty is associated with the numerous cuts which were applied in order to isolate the reaction of interest, i.e., $e p \to  e^\prime p^\prime \eta$  
To estimate the systematic uncertainty of a cut, the value of the cut was
varied from the standard cut position by a  step on each side by $\pm0.5\sigma$,
where $\sigma$ is the resolution of the corresponding variable. Thus, the resulting  cross sections and structure functions  were obtained at each of 4 cut values in addition to the standard cut of $\pm3\sigma$.

All cuts were varied independently,  such that at each cut iteration, for each distribution, the entire  analysis, including calculation of acceptances, cross sections, radiative corrections and structure functions was performed. Then, for each kinematic point, the  cross sections and structure functions were plotted as functions of cut variation and a linear fit was
performed. The slope parameter of the fit was assumed to be the systematic
uncertainty introduced by the particular cut at a given kinematic point. This procedure was
performed for  all  sources of kinematic uncertainties where it was applicable.
It was shown that this method of systematic uncertainty calculation
overestimates  the systematic uncertainty for bins with low statistics, but was retained.

The systematic uncertainty   associated with the variation of  the cross section within a kinematic  bin  at $Q^2$, $x_B$ and  $t$  
was estimated to be $\pm 1.3$\% by using our cross section model. 

\begin{figure}
\begin{centering} 
\includegraphics[width=\columnwidth]{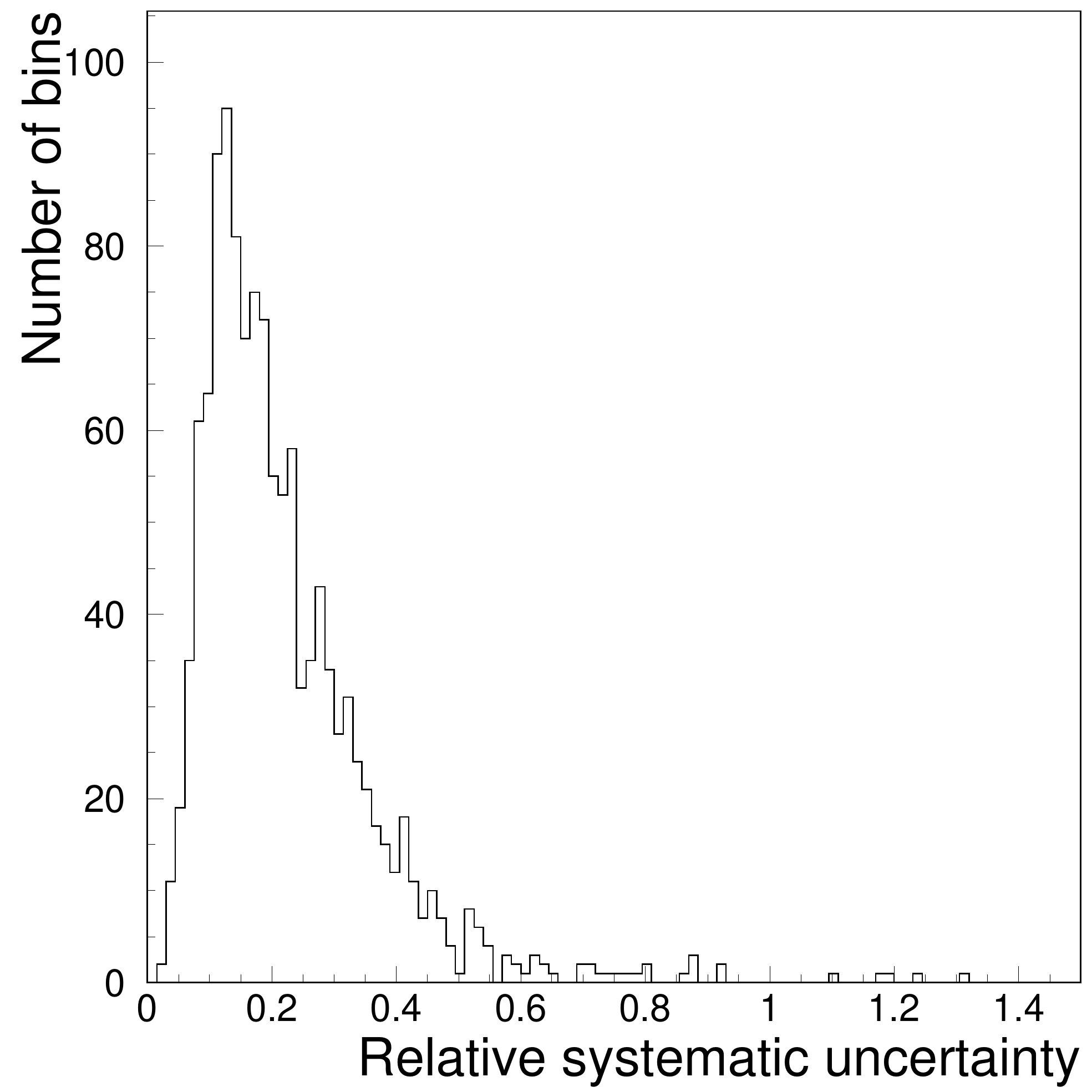}  
\caption{\label{fig:syst-tot} 
The relative systematic  uncertainties, $\delta\sigma_{sys}/\sigma$ of the fourfold differential cross section (see Eq. \ref{eq:sig_ep_eppippim}) for all kinematic points. These do not  include the overall normalization uncertainty, 
}
\end{centering} 
\end{figure}

To estimate the systematic uncertainty of the absolute normalization procedure, the normalization constant $\delta_{Norm}$ was obtained separately for electrons detected in each of the six sectors, resulting in a mean value of 87\%.  The sector-by-sector rms variation from the mean value  was used as an estimate of the systematic uncertainty on the mean. 
The distribution of total systematic uncertainty, excluding the uncertainty on absolute normalization is shown in  Fig.~\ref{fig:syst-tot}.
Table~\ref{table:syst-summary} contains a summary of the information on all of the sources of systematic uncertainty on the individual fourfold differential cross sections - \ $ \frac{d^4 \sigma_{ep \rightarrow e^\prime p^\prime \eta}}{dQ^2 dx_B dt d\phi_\eta}$ -\  that were
studied.

\begingroup
\squeezetable
\begin{table*}
\caption{Summary table of systematic uncertainties}
\centering
\begin{ruledtabular}
\begin{tabular}{lccc}
Source & Varies  & Average uncertainty & Average uncertainty   \\
                                 & by bin  & of the cross section  & of the structure function $\sigma_U$ \\
\hline
Target length & No & 0.2\% & 0.2\% \\
Electron fiducial cut & Yes &$\sim  6.4\%$ &$\sim  3.5\%$  \\
Proton fiducial cut & Yes &$\sim  4.1\%$ &$\sim  2.4\%$  \\
Cut on missing mass of the $e\gamma\gamma$ & Yes & $\sim3.9\%$ & $\sim0.7\%$\\
Cut on invariant mass of 2 photons & Yes & $\sim10.5\%$ & $\sim 9.0\%$ \\
Cut on missing energy of the $ep\gamma\gamma$ & Yes & $\sim 6.6\%$ & $\sim 4.1\%$ \\
Radiative corrections and cut on $M_X(ep)$ & Yes & $\sim 8.0\%$ &  $\sim 6.0\%$ \\
Absolute normalization & No & $4.1\%$ & $4.1\%$ \\
Luminosity calculation & No & $ < 1\%$ & $ < 1\%$ \\
Bin volume correction      & Yes& $\sim 1.3\%$ & $\sim 1.3\%$ \\
Cut on energy of photon detected in the EC & Yes & $\sim 3.1\%$& $\sim 2.5\%$ \\
\end{tabular}
\end{ruledtabular}
\label{table:syst-summary}
\end{table*} 
\endgroup

\section{Structure functions  }
The reduced cross sections can be expanded in terms of  structure functions as follows:

\begin{equation}
\begin{split}
&2\pi\frac{d^2\sigma}{dtd\phi_\eta} 
=\left( \frac{d\sigma_T}{dt}+\epsilon\frac{d\sigma_L}{dt}\right)+\\
&\epsilon \cos 2 \phi_\eta\frac{d\sigma_{TT}}{dt}
+ \sqrt{2\epsilon(1+\epsilon)}\cos \phi_\eta \frac{d\sigma_{LT}}{dt}, 
\end{split}
\label{eq:sigmaphidep}
\end{equation}

\noindent
from which the three combinations of structure functions, 

\begin{equation}
\frac{d\sigma_U}{dt}\equiv \frac{d\sigma_T}{dt}+\epsilon\frac{d\sigma_L}{dt},\  \  \frac{d\sigma_{TT}}{dt}\   \text{and}\ 
\frac{d\sigma_{LT}}{dt} 
\end{equation}

\noindent can be extracted by fitting the cross sections to the $\phi_\eta$ distribution in each bin of $(Q^2,x_B,t)$. As an example,  the  curve in Fig.~\ref{fig:cross-section} is a fit to $d^2\sigma/dtd\phi_\eta$ in terms of the coefficients of the $\cos \phi_\eta$ and $\cos 2\phi_\eta$ terms.
The physical significance of the structure functions is as follows.
\begin{itemize}
\item $d\sigma_L/dt$  is the sum of structure functions initiated by a longitudinal virtual photon, both with and without nucleon helicity-flip, i.e., respectively $\Delta \nu = \pm 1$ and $\Delta \nu = 0$;

\item $d\sigma_T/dt$ is the sum of structure functions initiated by  transverse virtual photons of positive and negative helicity ($\mu = \pm 1$), with and without nucleon helicity flip, respectively $\Delta \nu = \pm 1$ and  $0$;

\item $d\sigma_{LT}/dt$ corresponds to interferences involving products  of amplitudes for longitudinal and transverse photons;

\item $d\sigma_{TT}/dt$ corresponds to interferences involving products  of transverse positive and negative photon helicity amplitudes.

\end{itemize}

The structure functions for all kinematic bins are shown in Fig.~\ref{fig:sigma_U_TT_LT} and listed in Appendix~\ref{app_strfun}. The quoted statistical uncertainties on the structure functions were obtained in the fitting  procedure taking into account the statistical uncertainties on the individual cross section points.  The quoted systematic uncertainties are the  variations of the fitted structure functions due to variation of the cut parameters.

\begin{figure*}
\includegraphics[width=\textwidth]{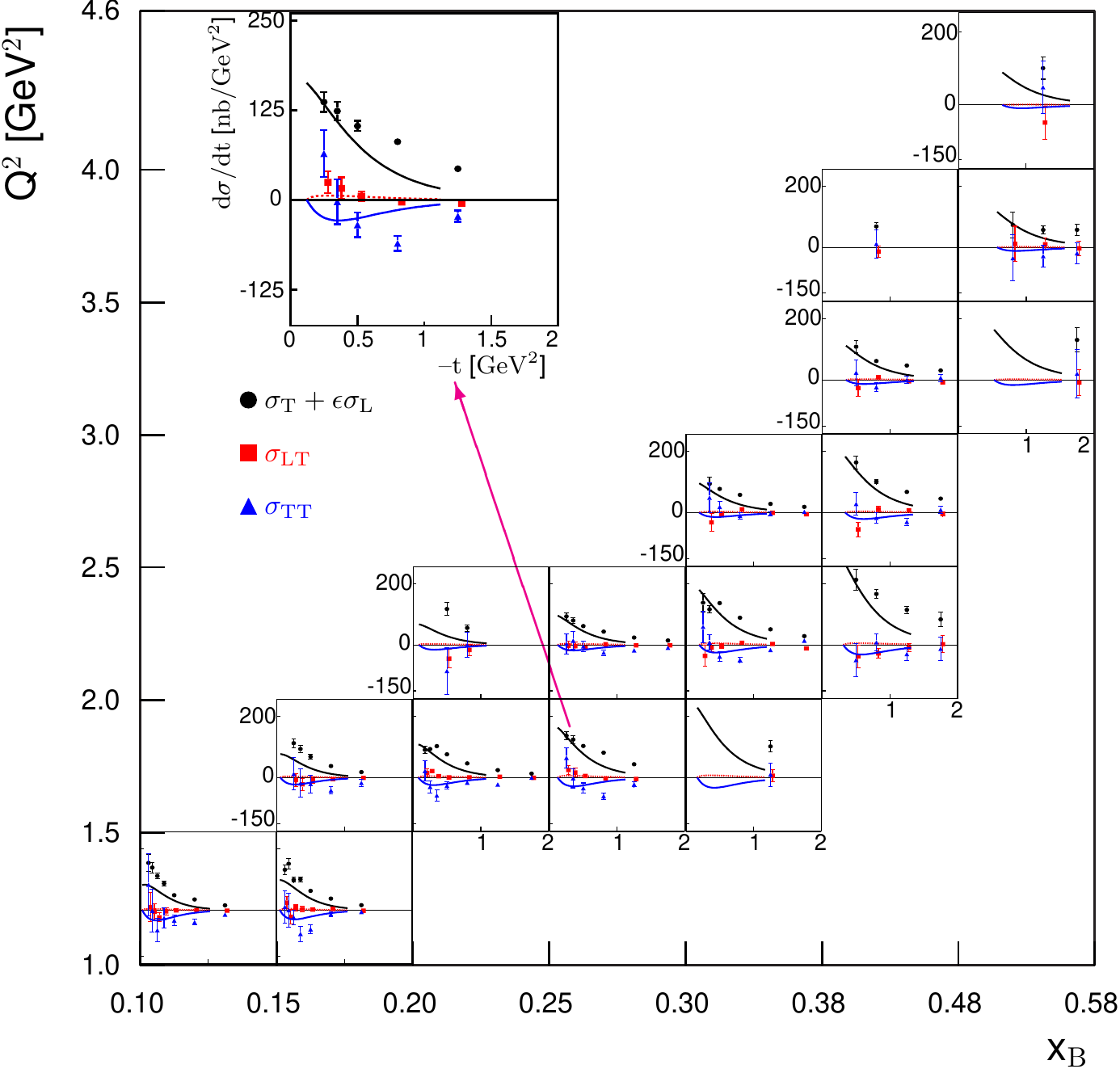}

\caption{The structure functions vs $t$ for the different $(Q^2,x_B)$ bins, extracted from the present experiment. 
Black circules: $d\sigma_U/dt$. Red squares: $d\sigma_{LT}/dt$. Blue triangles: $d\sigma_{TT}/dt$. 
The black, red and blue curves are the corresponding results of the  handbag based calculation of 
Ref.~\cite{Goloskokov:2011rd}.
The inset is an enlarged view of the bin with $x_B=0.17$ and $Q^2=1.87$ GeV$^2$.
The error bars are statistical only.
}
\label{fig:sigma_U_TT_LT}
\end{figure*}

 A number of observations can be made independently of the model predictions. The 
$d\sigma_{TT}/dt$ structure function is negative  and  is smaller in magnitude than unpolarized structure function 
 ($d\sigma_U/dt \equiv d\sigma_T/dt+\epsilon d\sigma_L/dt$).
  However, $d\sigma_{LT}/dt$ is significantly smaller  than $d\sigma_{TT}/dt$.
 This reinforces the conclusion that  the transverse photon amplitudes  are dominant at the present values of $Q^2$.

\begin{figure*}
\includegraphics[width=\textwidth]{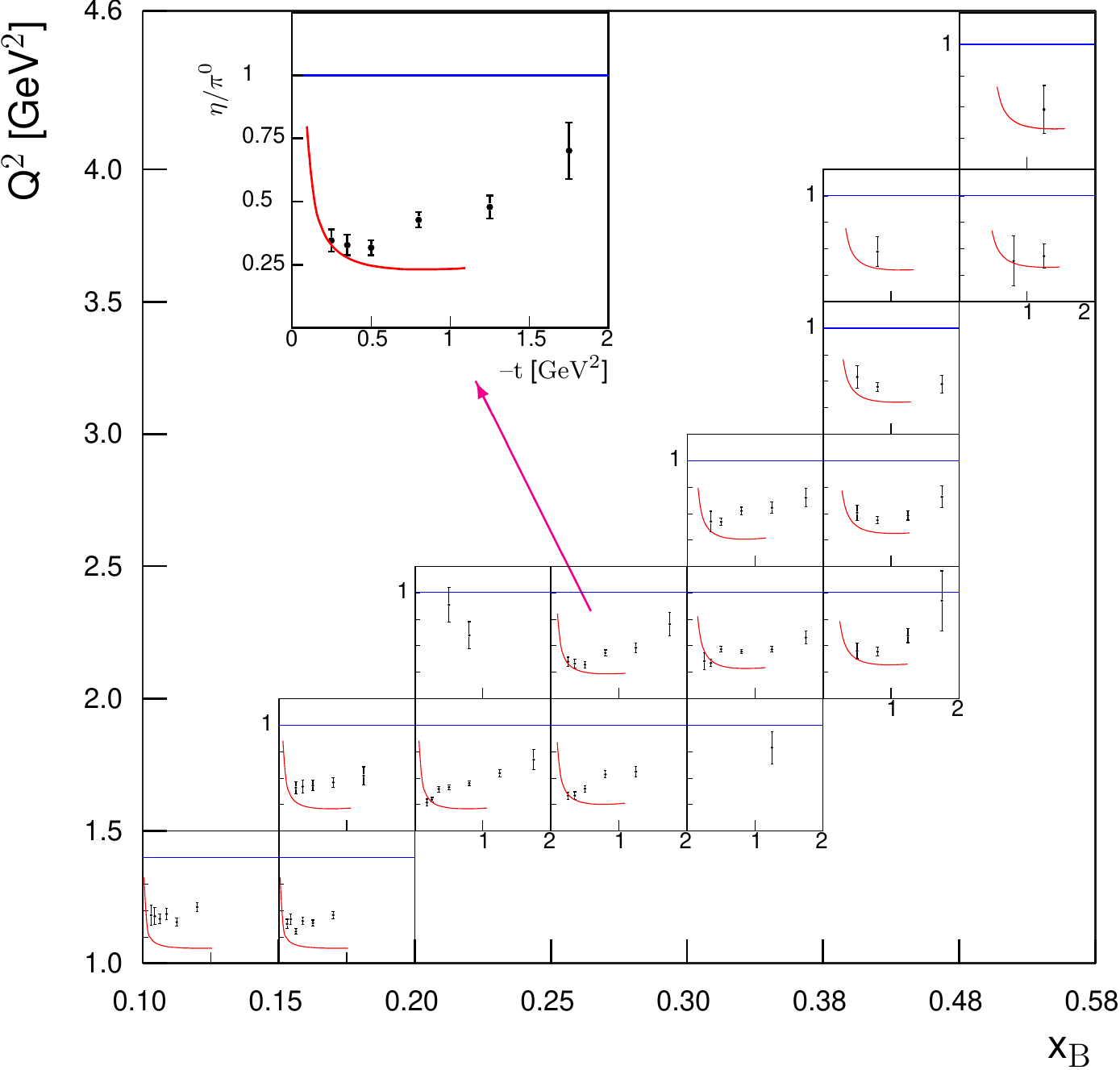}
\caption{
The ratio $R$ of the unpolarized structure functions for $\eta$ and $\pi^0$ extracted from the present experiment and Ref.~\cite{Bedlinskiy:2012be}, as functions of $t$ for $(Q^2,x_B)$ bins.   The leading order handbag calculations~\cite{Eides:1998ch} predict asymptotically $R \sim 1$. The curves are the result of a  handbag based calculation of Ref.~\cite{Goloskokov:2011rd}.
The inset is an enlarged view of the bin with $x_B=0.28$ and $Q^2=2.2$ GeV$^2$.
The error bars are statistical only.
}
\label{fig:ratio}
\end{figure*}

The ratio {\it R} of the unpolarized cross sections for $\eta$ and $\pi^0$  for all kinematic bins is shown in Fig.~\ref{fig:ratio}.  The ratio {\it R} is seen to be significantly less than 1, whereas the leading order handbag calculations ~\cite{Eides:1998ch} predict asymptotically $R \sim 1$. However, the observed value of {\it R}, typically about fifty percent, is greater than that predicted by the model of Ref.~\cite{Goloskokov:2011rd}.

\section{$t$- slopes}

After the structure functions were obtained, fits
were made to extract the $t$-dependence of $\sigma_U$ for different values $x_B$ and $Q^2$.
For each given $x_B$ and $Q^2$ we fit this structure function with an exponential function: $$\frac{d\sigma_U}{dt}=Ae^{Bt}.$$

\begin{figure}
\includegraphics[width=\columnwidth]{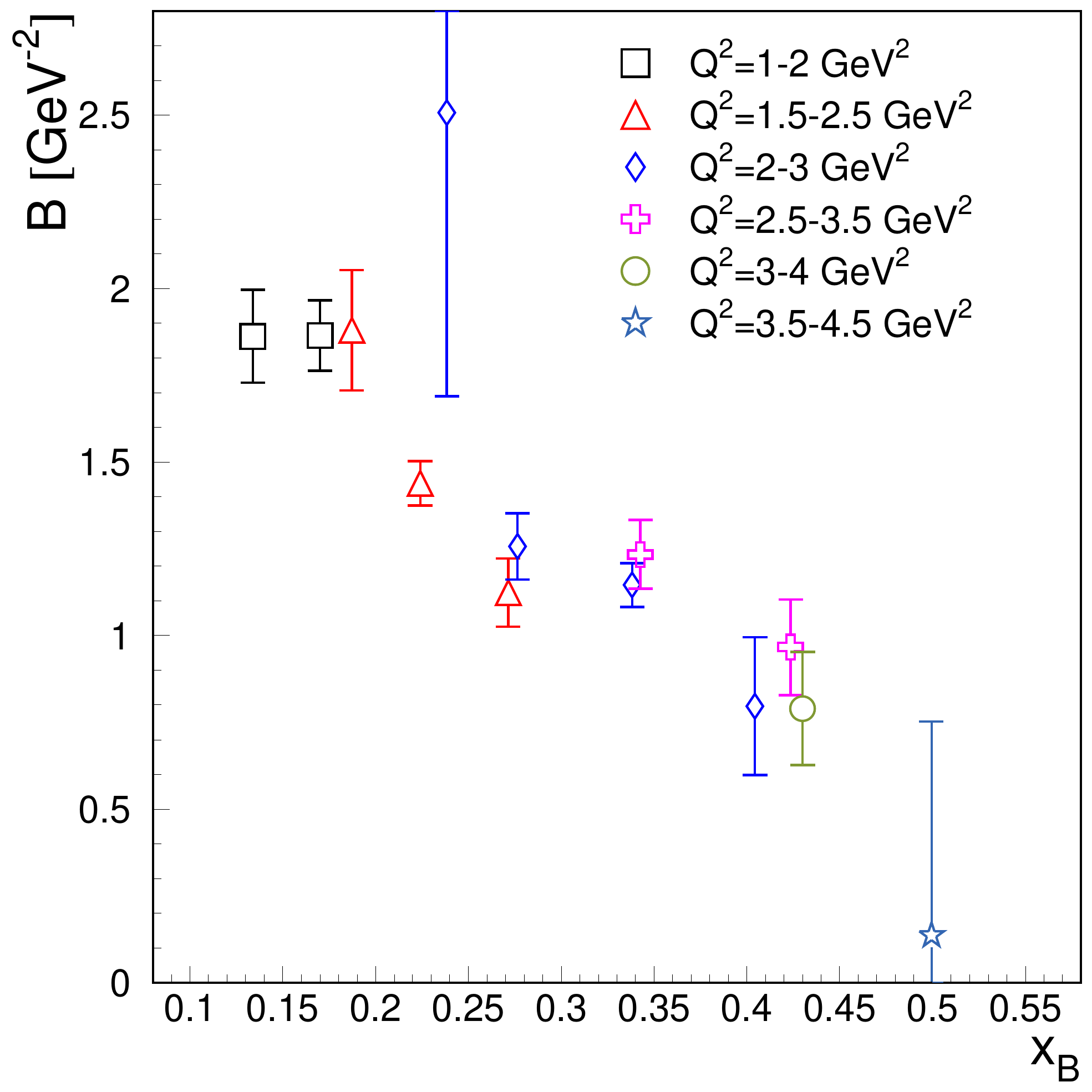}
\caption{Slope parameters $B$ for different $x_B$ and $Q^2$ bins.
The error bars are statistical only.}		
\label{fig:t-dep}
\end{figure}

Figure ~\ref{fig:t-dep} shows the slope parameter $B$ as a function of $x_B$ for different values of $Q^2$. 
The data appear to exhibit a decrease in slope parameter  with increasing $x_B$. However, the $Q^2-x_B$ correlation in the CLAS acceptance (see Fig.~\ref{fig:kin_cuts}) does not permit one to make a definite conclusion about the  $Q^2$ dependences of the slope parameter for fixed $x_B$. What one can say is that at high $Q^2$ and high $x_B$   the slope parameter appears to be smaller than for the lowest values of these variables.  
The $B$ parameter in the exponential determines the  width of the transverse momentum distribution of the emerging protons, which, by  a Fourier  transform, is inversely related to the transverse size of the interaction region.  From the point of view of the handbag picture, it is inversely related to  the  mean  transverse radius of the separation  between the active quark and the center of momentum of the spectators 
(see Ref.~\cite{Burkardt:2007sc}).  Thus the data implies that the separation is  larger at the lowest $x_B$ and $Q^2$ and becomes smaller for increasing $x_B$ and $Q^2$, as it must.  This is consistent with the results for $\pi^0$ electoproduction~\cite{Bedlinskiy:2014tvi}.

\section{Comparisons with Theoretical Handbag Models}

Figure ~\ref{fig:sigma_U_TT_LT} shows the experimental structure functions  for  bins of $Q^2$ and $x_B$.
The results of the GPD-based model of Goloskokov and Kroll \cite{Goloskokov:2011rd}  are  superimposed in Fig.~\ref{fig:sigma_U_TT_LT}. 
From these plots we  conclude that the GPD-based theoretical model generally describes the CLAS data in the kinematical region of this experiment, although there are systematic discrepancies. For example, the theoretical model appears to underestimate  $d\sigma_U/dt$ in most kinematic bins. 

According to GK,  
the primary contributing GPDs in meson production for transverse photons are  $H_T$, which characterizes the quark distributions involved in nucleon helicity-flip, and   $\bar E_T \ ( = 2\widetilde H_T + E_T), $ which characterizes the quark distributions  involved in  nucleon  helicity-non-flip 
processes~\cite{Diehl:2005jf, Gockeler:2006zu}. 
As a reminder, in both cases the active quark undergoes a helicity-flip. 
 The GPD $\bar E_T$  is related to the spatial density  of transversely  polarized quarks in an unpolarized nucleon~\cite{Gockeler:2006zu}.

\begin{figure*}
\centering
\includegraphics[width=13cm]{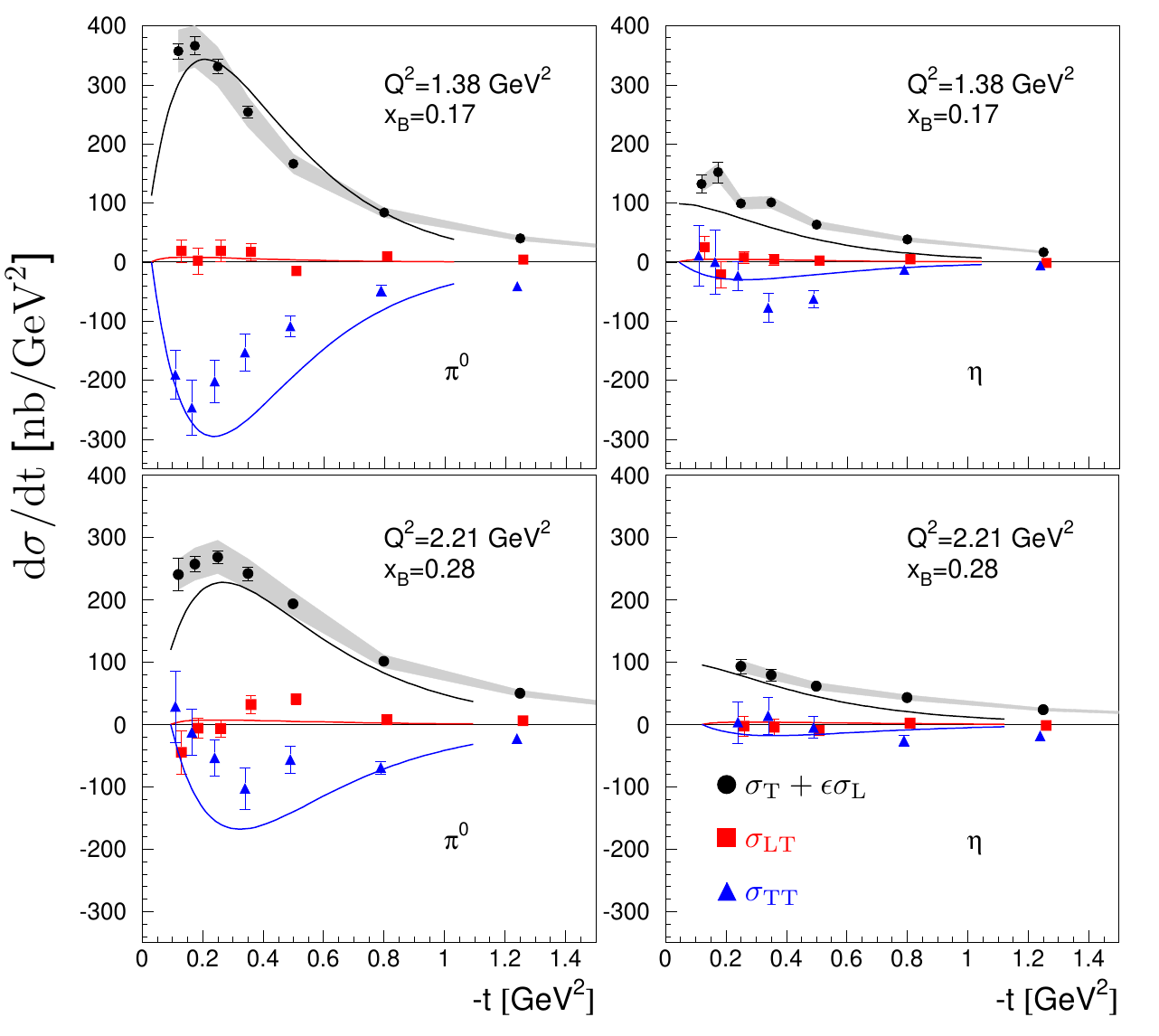}
\caption{
(Color online) 
The extracted structure functions vs $t$ for 
the $\pi^0$ (left column) 
~\cite{Agashe:2014kda} 
and $\eta$ (right column). The top row presents data for the kinematic point  ($Q^2=$1.38~GeV$^2$,$x_B$=0.17)
and bottom row for the kinematic point ($Q^2=$2.21~GeV$^2$,$x_B$=0.28). The data for the $\eta$ is identical to that shown in Fig. \ref{fig:sigma_U_TT_LT}, with the vertical  axis  rescaled to highlight the difference in the magnitude of the cross sections for $\pi^0$ and $\eta$ electroproduction. 
The data and curves are as follows:  
black circles - $d\sigma_U/dt =d\sigma_T/dt +\epsilon d\sigma_L/dt$,
blue triangles - $d\sigma_{TT}/dt$,
red  squares - $d\sigma_{LT}/dt$.
The error bars are statistical only. The gray bands are our estimates of the absolute normalization systematic uncertainties on $d\sigma_U/dt $.  The curves are theoretical predictions produced  with the models of 
Ref.~\cite{Goloskokov:2011rd}.}
\label{fig:compar}
\end{figure*} 

Ref.~\cite{Goloskokov:2011rd}  obtains the following relations:
\noindent 

\begin{equation}
\begin{split}
\frac{d\sigma_{T}}{dt} =& \frac{4\pi\alpha}{2k^\prime}\frac{\mu_\eta^2}{Q^8} 
                       \Bigl[ \left(1-\xi^2\right) \left|\GPDHT\right|^2 - \Bigl. \\
                                    & \Bigr.                \frac{t'}{8m^2} \left|\GPDETbar\right|^2\Bigr],
\end{split}
\label{sigmat}
\end{equation}

\begin{align}\label{sigmatt}
\frac{d\sigma_{TT}}{dt} = \frac{4\pi\alpha}{k^\prime}\frac{\mu_\eta^2}{Q^8}\frac{t'}{16m^2}\left|\GPDETbar\right|^2.
\end{align}

\noindent Here $\kappa^\prime(Q^2,x_B)$ is a phase space factor, $t^\prime =t-t_{min}$,
 and the brackets $\langle  H_T \rangle$ and $\langle \bar E_T \rangle$ are the Generalized Form Factors (GFFs) that denote
the convolution of the elementary process with the GPDs   $H_T$ and $\bar E_T$ (see Fig.~\ref{fig:handbag-pi0}).

Note that for the case of nucleon helicity-non-flip, characterized by  the GPD $\bar E_T$, overall helicity from the initial to the final state is not conserved. However, angular momentum  is 
conserved \emph{-} the difference being absorbed by the orbital motion of the scattered 
$\eta-N$ pair.  This accounts for the additional $t^\prime$ factor multiplying the $\bar E_T$ terms in Eqs. \ref{sigmat} and \ref{sigmatt}.

As in the case of $\pi^0$ electroproduction, the contribution of $\sigma_L$  accounts for only a small fraction  of the unseparated  structure functions $d\sigma_U/dt (\equiv d\sigma_T/dt+ \epsilon d\sigma_L/dt)$ in the kinematic regime under investigation. This is because the contributions from $\tilde H$  and $\tilde E$ - the GPDs that are responsible for the leading-twist structure function $\sigma_L$ -  are  relatively small compared with the contributions from $\bar E_T$  and $H_T$ (although not quite as small for $\eta$ production as compared with $\pi^0$ production), which contribute to $d\sigma_T/dt$ and $d\sigma_{TT}/dt$. 
The extracted structure functions  at selected values  of $Q^2$ and $x_B$ for 
the $\pi^0$ (left column) and $\eta$ (right column) are shown in Fig.~\ref{fig:compar} side-by-side. The top row represents data for the kinematic point  ($Q^2=$1.38~GeV$^2$, $x_B$=0.17)
and the bottom row for the kinematic point ($Q^2=$2.21~GeV$^2$, $x_B$=0.28).
The unpolarized structure function $d\sigma_U/dt$ for $\eta$ production is   significantly smaller than that for $\pi^0$ for all measured kinematic intervals of $Q^2, x_B$ and $t$. This is in contradiction to the leading order calculation ~\cite{Eides:1998ch} with $d\sigma_L/dt$ dominance,  where the ratio is expected to be on the order of unity.  In the present case,  $\bar E_T$ is significantly larger than  $H_T$.
 The curves in Fig.~\ref{fig:sigma_U_TT_LT} and \ref{fig:compar} are obtained by GK~\cite{Goloskokov:2011rd}. For the GPDs,  their parametrization was guided by the lattice calculation results of Ref.~\cite{Gockeler:2006zu}.

The  relative importance of $\bar E_T$  and $H_T$ can be understood by considering their  composition in terms of their valence quark flavors and  GPDs. Following GK, the $ \pi^0$ and $\eta$ GPDs in terms of valence quark GPDs may be expressed as follows.
\noindent  For $ \pi^0:$
\begin{equation}    
\begin{split}
 H_T^{\pi^0} =(e_u H_T^u - e_d  H_T^d) /{\sqrt{2}} , \\
 \bar E_T^{\pi^0} =(e_u \bar E_T^{u} - e_d  \bar E_T^{d}) / {\sqrt{2}},
\label{eq:pi0}
\end{split}
\end{equation}
\noindent where $e_u=1/3$ and $e_d =-2/3$. 

For $\eta$, assuming the valence structure of the $\eta$ is purely a member of the SU(3) octet, i.e., $\eta = \eta_8$,
and there is no contribution from strange quarks is
\begin{equation}   \label{eq:eta}
\begin{split}
 H_T^{\eta} =(e_u H_T^u +e_d  H_T^d) /{\sqrt{6}} ,\\
 \bar E_T^{\eta} =(e_u \bar E_T^{u} + e_d  \bar E_T^{d}) / {\sqrt{6}}.
\end{split}
\end{equation}
\noindent
In the model of GK, the sign of  $H_T^u$ is positive, while the sign of $H_T^d$ is negative, but the signs of $\bar E_T^{u}$ and 
$\bar E_T^{d}$ are both positive.  Thus, for  $\pi^0$,  taking into account the sign of $e_u$ and $e_d$, the up and down quarks enhance $\bar E_T^{\pi^0}$ and diminish $H_{T}^{\pi^0}$.
The opposite effect occurs for  $\eta$ mesons. 
By combining the  $\eta$ and $\pi^0$ data, and Eqs. \ref{eq:pi0} and \ref{eq:eta} above, one can estimate the GPDs of the individual valence quark flavors in the framework of the dominance of the transversity GPDs. This is currently underway \cite{Kubarovsky:2016} and will be presented later.

We further note the following  features:
for $\eta$ production the model of GK appears to underestimate the magnitude of $d\sigma_{U}/dt$, whereas for $\pi^0$ electroproduction the  theoretical calculation of $d\sigma_{U}/dt$ more closely agrees with the data. Thus, one is led to the hypothesis that possibly $H_T$  is underestimated for $\eta$ electroproduction. Increasing $H_T$ will increase $d\sigma_{T}/dt$ and, therefore, $d\sigma_{U}/dt$, while not affecting $d\sigma_{TT}/dt$.

Referring again to Fig. \ref{fig:ratio}, which shows the ratio of $d\sigma_U/dt$ for $\eta$ and $\pi^0$, the experimental value of this ratio is systematically higher than the theoretical prediction, which is related to the underestimation of the $\eta$ cross section.

\section{Conclusion}

Differential cross sections of exclusive $\eta$ electroproduction were  obtained in the few-GeV region in  bins of $Q^2, x_B$, $t$ and $\phi_\eta$.  
Virtual photon structure functions  $d\sigma_U/dt=d(\sigma_T+\epsilon\sigma_L)/dt$, $d\sigma_{TT}/dt$ and $d\sigma_{LT}/dt$ were  extracted. It is found that $d\sigma_U/dt$  is larger in magnitude than $d\sigma_{TT}/dt$,  while $d\sigma_{LT}/dt$ is significantly smaller than $d\sigma_{TT}/dt$. The exclusive cross sections and structure functions are typically more than a factor of two smaller than for previously measured $\pi^0$ electroproduction for similar kinematic intervals. 
It appears that some of these differences can be roughly understood from GPD-models in terms of the quark composition of $\pi^0$ and $\eta$ mesons.
The cross section ratios of $\eta$ to $\pi^0$  appear to agree with the handbag calculations at low $|t|$, but show  significant deviations with increasing $|t|$.

Within the handbag interpretation, there are  theoretical calculations \cite{ Goloskokov:2011rd}, which were earlier found to describe  $\pi^0$ electroproduction~\cite{Bedlinskiy:2014tvi} quite well.  The result  of the calculations confirmed that the  measured unseparated cross sections are much larger than expected from leading-twist handbag calculations, which are dominated by longitudinal photons. For the present case,  the  same conclusion can be made in an almost model independent way by noting that the structure functions $d\sigma_U/dt$ and $d\sigma_{TT}/dt$ are  significantly larger than $d\sigma_{LT}/dt$.

To make significant improvement in interpretation, higher statistical precision data, as well as $L\textrm{-}T$ separation and polarization measurements over the entire range of kinematic variables are necessary. Such experiments are planned for the  Jefferson Lab operations at 12 GeV.

\begin{acknowledgments}
We thank  the staff of the Accelerator and Physics Divisions at Jefferson Lab for making the experiment possible. We also thank 
G.~Goldstein, 
S.~Goloskokov, 
P.~Kroll, 
J. M.~Laget, 
S.~Liuti and 
A.~Radyushkin 
for many informative discussions,  and clarifications of their work, and for making available the results of their calculations. 
This work was supported in part by 
the U.S. Department of Energy (DOE) and National Science Foundation (NSF), 
the French Centre National de la Recherche Scientifique (CNRS) and Commissariat  \`a l'Energie Atomique (CEA), the French-American Cultural Exchange (FACE),
the Italian Istituto Nazionale di Fisica Nucleare (INFN), 
the Chilean Comisi\'on Nacional de Investigaci\'on Cient\'ifica y Tecnol\'ogica (CONICYT),
the National Research Foundation of Korea, 
and the UK Science and Technology Facilities
Council (STFC).
The Jefferson Science Associates (JSA) operates the Thomas Jefferson National Accelerator Facility for 
the United States Department of Energy under contract DE-AC05-06OR23177. 
\end{acknowledgments}

\appendix
\begin{widetext}
\section{Structure Functions}
\label{app_strfun}
The structure functions are presented in Table ~\ref{strfun_table}. The first error is statistical uncertainty and the second  is the systematic uncertainty.
\squeezetable
\setlength\LTleft{0pt}
\setlength\LTright{0pt}
\begin{longtable}{ccc@{\extracolsep{1cm}}d@{\extracolsep{0pt}}cdcd@{\extracolsep{1cm}}d@{\extracolsep{0pt}}cdcd@{\extracolsep{1cm}}d@{\extracolsep{0pt}}cdcd}
\caption{Structure Functions} \label{strfun_table} \\
\hline 
\hline
\multicolumn{1}{c}{\textbf{$Q^2$,}} &
\multicolumn{1}{c}{\textbf{$x_B$}} &
\multicolumn{1}{c}{\textbf{$-t$,}} &
\multicolumn{5}{c}{\textbf{$\frac{d\sigma_T}{dt}+\epsilon\frac{d\sigma_L}{dt}$,}} &
\multicolumn{5}{c}{\textbf{$\frac{d\sigma_{LT}}{dt}$,}} &
\multicolumn{5}{c}{\textbf{$\frac{d\sigma_{TT}}{dt}$,}} \\
\multicolumn{1}{c}{\textbf{$GeV^2$}} &
\multicolumn{1}{c}{\textbf{}} &
\multicolumn{1}{c}{\textbf{$GeV^2$}} &
\multicolumn{5}{c}{\textbf{$nb/GeV^2$}} &
\multicolumn{5}{c}{\textbf{$nb/GeV^2$}} &
\multicolumn{5}{c}{\textbf{$nb/GeV^2$}}
\\\hline
\endfirsthead
\hline 
\hline
\multicolumn{1}{c}{\textbf{$Q^2$,}} &
\multicolumn{1}{c}{\textbf{$x_B$}} &
\multicolumn{1}{c}{\textbf{$-t$,}} &
\multicolumn{5}{c}{\textbf{$\frac{d\sigma_T}{dt}+\epsilon\frac{d\sigma_L}{dt}$,}} &
\multicolumn{5}{c}{\textbf{$\frac{d\sigma_{LT}}{dt}$,}} &
\multicolumn{5}{c}{\textbf{$\frac{d\sigma_{TT}}{dt}$,}} \\
\multicolumn{1}{c}{\textbf{$GeV^2$}} &
\multicolumn{1}{c}{\textbf{}} &
\multicolumn{1}{c}{\textbf{$GeV^2$}} &
\multicolumn{5}{c}{\textbf{$nb/GeV^2$}} &
\multicolumn{5}{c}{\textbf{$nb/GeV^2$}} &
\multicolumn{5}{c}{\textbf{$nb/GeV^2$}}
\\\hline
\endhead 
\hline
\endfoot
\hline \hline
\endlastfoot
1.17 & 0.134 & 0.12 & 159.3 &	$\pm$ & 27.7 &	$\pm$ & 22.3 & 8.2 &	$\pm$ & 49.3 &	$\pm$ & 33.2 & 88.4 &	$\pm$ & 104.2 &	$\pm$ & 126.4\\
1.17 & 0.134 & 0.17 & 144.7 &	$\pm$ & 18.0 &	$\pm$ & 16.2 & 2.2 &	$\pm$ & 26.4 &	$\pm$ & 20.2 & -4.3 &	$\pm$ & 73.1 &	$\pm$ & 189.0\\
1.17 & 0.134 & 0.25 & 117.3 &	$\pm$ & 10.3 &	$\pm$ & 10.7 & -22.0 &	$\pm$ & 14.9 &	$\pm$ & 9.9 & -71.6 &	$\pm$ & 40.2 &	$\pm$ & 29.1\\
1.17 & 0.134 & 0.35 & 94.0 &	$\pm$ & 8.8 &	$\pm$ & 3.6 & -1.3 &	$\pm$ & 12.7 &	$\pm$ & 4.2 & -29.7 &	$\pm$ & 35.7 &	$\pm$ & 9.0\\
1.17 & 0.134 & 0.50 & 51.1 &	$\pm$ & 4.3 &	$\pm$ & 5.9 & 1.8 &	$\pm$ & 6.0 &	$\pm$ & 4.4 & -34.1 &	$\pm$ & 18.2 &	$\pm$ & 10.0\\
1.17 & 0.134 & 0.80 & 36.3 &	$\pm$ & 2.5 &	$\pm$ & 1.6 & 1.1 &	$\pm$ & 3.0 &	$\pm$ & 5.6 & -40.6 &	$\pm$ & 9.5 &	$\pm$ & 13.3\\
1.17 & 0.134 & 1.25 & 16.2 &	$\pm$ & 1.7 &	$\pm$ & 1.8 & -1.2 &	$\pm$ & 2.3 &	$\pm$ & 3.0 & -13.7 &	$\pm$ & 6.2 &	$\pm$ & 5.0\\
1.39 & 0.170 & 0.12 & 134.1 &	$\pm$ & 15.5 &	$\pm$ & 21.7 & 26.2 &	$\pm$ & 19.8 &	$\pm$ & 14.2 & 15.2 &	$\pm$ & 52.7 &	$\pm$ & 27.5\\
1.39 & 0.170 & 0.17 & 156.4 &	$\pm$ & 18.2 &	$\pm$ & 21.9 & -18.1 &	$\pm$ & 23.3 &	$\pm$ & 28.7 & -0.4 &	$\pm$ & 56.5 &	$\pm$ & 8.0\\
1.39 & 0.170 & 0.25 & 101.8 &	$\pm$ & 8.0 &	$\pm$ & 7.9 & 10.6 &	$\pm$ & 10.0 &	$\pm$ & 6.4 & -22.9 &	$\pm$ & 25.1 &	$\pm$ & 26.2\\
1.39 & 0.170 & 0.35 & 104.6 &	$\pm$ & 8.0 &	$\pm$ & 6.3 & 7.6 &	$\pm$ & 9.3 &	$\pm$ & 9.2 & -80.1 &	$\pm$ & 25.3 &	$\pm$ & 15.4\\
1.39 & 0.170 & 0.50 & 65.3 &	$\pm$ & 4.5 &	$\pm$ & 2.7 & 4.3 &	$\pm$ & 5.0 &	$\pm$ & 3.1 & -64.3 &	$\pm$ & 14.9 &	$\pm$ & 16.7\\
1.39 & 0.170 & 0.80 & 39.0 &	$\pm$ & 2.4 &	$\pm$ & 2.6 & 5.7 &	$\pm$ & 2.8 &	$\pm$ & 3.3 & -11.9 &	$\pm$ & 8.0 &	$\pm$ & 4.5\\
1.39 & 0.170 & 1.25 & 16.9 &	$\pm$ & 1.5 &	$\pm$ & 2.1 & -1.7 &	$\pm$ & 1.9 &	$\pm$ & 1.1 & -6.0 &	$\pm$ & 5.2 &	$\pm$ & 2.9\\
1.62 & 0.187 & 0.25 & 117.1 &	$\pm$ & 14.6 &	$\pm$ & 11.6 & -6.0 &	$\pm$ & 22.0 &	$\pm$ & 13.4 & 11.3 &	$\pm$ & 54.6 &	$\pm$ & 32.0\\
1.62 & 0.187 & 0.35 & 98.4 &	$\pm$ & 13.2 &	$\pm$ & 9.0 & -20.3 &	$\pm$ & 20.4 &	$\pm$ & 6.8 & -22.0 &	$\pm$ & 48.6 &	$\pm$ & 49.5\\
1.62 & 0.187 & 0.50 & 71.0 &	$\pm$ & 7.6 &	$\pm$ & 3.6 & -5.7 &	$\pm$ & 10.7 &	$\pm$ & 6.9 & -22.7 &	$\pm$ & 30.7 &	$\pm$ & 37.5\\
1.62 & 0.187 & 0.80 & 38.5 &	$\pm$ & 3.3 &	$\pm$ & 1.7 & -4.3 &	$\pm$ & 4.4 &	$\pm$ & 2.1 & -43.0 &	$\pm$ & 12.4 &	$\pm$ & 8.7\\
1.62 & 0.187 & 1.25 & 18.3 &	$\pm$ & 2.7 &	$\pm$ & 2.2 & -1.2 &	$\pm$ & 3.8 &	$\pm$ & 1.6 & -15.9 &	$\pm$ & 11.5 &	$\pm$ & 5.8\\
1.77 & 0.224 & 0.18 & 93.3 &	$\pm$ & 11.4 &	$\pm$ & 12.0 & 16.9 &	$\pm$ & 14.7 &	$\pm$ & 11.9 & 22.1 &	$\pm$ & 33.7 &	$\pm$ & 29.9\\
1.77 & 0.224 & 0.25 & 96.4 &	$\pm$ & 6.4 &	$\pm$ & 6.7 & 23.9 &	$\pm$ & 7.2 &	$\pm$ & 6.1 & -30.0 &	$\pm$ & 20.0 &	$\pm$ & 14.9\\
1.77 & 0.224 & 0.35 & 105.0 &	$\pm$ & 6.6 &	$\pm$ & 4.1 & 7.7 &	$\pm$ & 7.0 &	$\pm$ & 6.1 & -60.1 &	$\pm$ & 19.3 &	$\pm$ & 13.5\\
1.77 & 0.224 & 0.50 & 77.9 &	$\pm$ & 4.0 &	$\pm$ & 4.2 & 2.8 &	$\pm$ & 4.4 &	$\pm$ & 3.3 & -25.4 &	$\pm$ & 11.7 &	$\pm$ & 17.3\\
1.77 & 0.224 & 0.80 & 46.9 &	$\pm$ & 2.2 &	$\pm$ & 3.2 & 2.1 &	$\pm$ & 2.4 &	$\pm$ & 2.1 & -15.5 &	$\pm$ & 6.5 &	$\pm$ & 6.6\\
1.77 & 0.224 & 1.25 & 24.5 &	$\pm$ & 1.5 &	$\pm$ & 1.8 & 3.0 &	$\pm$ & 1.5 &	$\pm$ & 1.8 & -22.5 &	$\pm$ & 4.2 &	$\pm$ & 2.7\\
1.77 & 0.224 & 1.75 & 12.9 &	$\pm$ & 1.7 &	$\pm$ & 1.5 & -0.9 &	$\pm$ & 2.1 &	$\pm$ & 1.8 & -0.5 &	$\pm$ & 4.9 &	$\pm$ & 4.5\\
1.88 & 0.271 & 0.25 & 137.5 &	$\pm$ & 13.8 &	$\pm$ & 27.9 & 27.4 &	$\pm$ & 15.4 &	$\pm$ & 19.3 & 62.5 &	$\pm$ & 33.0 &	$\pm$ & 46.8\\
1.88 & 0.272 & 0.35 & 125.9 &	$\pm$ & 13.3 &	$\pm$ & 11.5 & 18.9 &	$\pm$ & 15.3 &	$\pm$ & 14.7 & -1.1 &	$\pm$ & 31.3 &	$\pm$ & 78.2\\
1.88 & 0.271 & 0.50 & 104.0 &	$\pm$ & 7.1 &	$\pm$ & 3.7 & 6.5 &	$\pm$ & 6.7 &	$\pm$ & 6.4 & -34.3 &	$\pm$ & 17.2 &	$\pm$ & 31.1\\
1.88 & 0.272 & 0.80 & 81.9 &	$\pm$ & 4.7 &	$\pm$ & 5.1 & -2.3 &	$\pm$ & 4.0 &	$\pm$ & 3.0 & -60.5 &	$\pm$ & 10.5 &	$\pm$ & 10.5\\
1.88 & 0.272 & 1.25 & 43.6 &	$\pm$ & 3.4 &	$\pm$ & 5.6 & -4.0 &	$\pm$ & 3.4 &	$\pm$ & 4.4 & -23.2 &	$\pm$ & 7.8 &	$\pm$ & 7.0\\
1.95 & 0.313 & 1.25 & 100.9 &	$\pm$ & 18.2 &	$\pm$ & 10.3 & 6.9 &	$\pm$ & 18.6 &	$\pm$ & 18.9 & 9.5 &	$\pm$ & 38.4 &	$\pm$ & 34.7\\
2.11 & 0.238 & 0.50 & 121.5 &	$\pm$ & 21.1 &	$\pm$ & 10.5 & -42.3 &	$\pm$ & 29.7 &	$\pm$ & 8.6 & -96.2 &	$\pm$ & 78.9 &	$\pm$ & 16.2\\
2.11 & 0.238 & 0.80 & 55.8 &	$\pm$ & 10.6 &	$\pm$ & 6.6 & -14.2 &	$\pm$ & 18.4 &	$\pm$ & 4.0 & -1.4 &	$\pm$ & 41.5 &	$\pm$ & 83.4\\
2.24 & 0.276 & 0.25 & 97.0 &	$\pm$ & 11.6 &	$\pm$ & 10.9 & -1.0 &	$\pm$ & 16.7 &	$\pm$ & 20.1 & 2.0 &	$\pm$ & 34.5 &	$\pm$ & 24.7\\
2.24 & 0.276 & 0.35 & 80.8 &	$\pm$ & 9.3 &	$\pm$ & 5.8 & -2.0 &	$\pm$ & 12.9 &	$\pm$ & 4.7 & 15.4 &	$\pm$ & 29.5 &	$\pm$ & 15.8\\
2.24 & 0.276 & 0.50 & 62.5 &	$\pm$ & 5.3 &	$\pm$ & 7.3 & -7.8 &	$\pm$ & 7.1 &	$\pm$ & 5.3 & -5.3 &	$\pm$ & 18.0 &	$\pm$ & 25.0\\
2.24 & 0.276 & 0.80 & 44.1 &	$\pm$ & 2.8 &	$\pm$ & 2.3 & 3.4 &	$\pm$ & 3.3 &	$\pm$ & 2.1 & -25.0 &	$\pm$ & 9.1 &	$\pm$ & 4.7\\
2.24 & 0.276 & 1.25 & 24.2 &	$\pm$ & 2.1 &	$\pm$ & 2.4 & -1.5 &	$\pm$ & 2.8 &	$\pm$ & 2.3 & -17.4 &	$\pm$ & 6.4 &	$\pm$ & 4.3\\
2.24 & 0.276 & 1.75 & 14.7 &	$\pm$ & 2.1 &	$\pm$ & 2.4 & -1.3 &	$\pm$ & 2.5 &	$\pm$ & 2.5 & -9.8 &	$\pm$ & 6.0 &	$\pm$ & 5.7\\
2.26 & 0.335 & 0.25 & 142.4 &	$\pm$ & 31.9 &	$\pm$ & 41.2 & -35.5 &	$\pm$ & 35.4 &	$\pm$ & 49.9 & 61.6 &	$\pm$ & 53.2 &	$\pm$ & 72.7\\
2.26 & 0.338 & 0.35 & 116.8 &	$\pm$ & 11.7 &	$\pm$ & 7.0 & -7.9 &	$\pm$ & 13.2 &	$\pm$ & 12.2 & 6.4 &	$\pm$ & 26.3 &	$\pm$ & 40.2\\
2.26 & 0.338 & 0.50 & 137.8 &	$\pm$ & 6.7 &	$\pm$ & 7.7 & -1.9 &	$\pm$ & 7.1 &	$\pm$ & 6.4 & -38.1 &	$\pm$ & 15.6 &	$\pm$ & 4.2\\
2.26 & 0.338 & 0.80 & 88.8 &	$\pm$ & 3.6 &	$\pm$ & 3.8 & 8.1 &	$\pm$ & 3.3 &	$\pm$ & 3.8 & -49.6 &	$\pm$ & 7.9 &	$\pm$ & 6.7\\
2.26 & 0.338 & 1.25 & 51.2 &	$\pm$ & 2.7 &	$\pm$ & 5.5 & 3.1 &	$\pm$ & 2.8 &	$\pm$ & 6.5 & -16.4 &	$\pm$ & 6.1 &	$\pm$ & 10.6\\
2.26 & 0.338 & 1.75 & 28.5 &	$\pm$ & 2.9 &	$\pm$ & 4.4 & -11.4 &	$\pm$ & 3.1 &	$\pm$ & 6.0 & 13.7 &	$\pm$ & 5.1 &	$\pm$ & 4.6\\
2.35 & 0.404 & 0.50 & 215.1 &	$\pm$ & 34.0 &	$\pm$ & 19.6 & -38.8 &	$\pm$ & 37.4 &	$\pm$ & 28.9 & -48.3 &	$\pm$ & 54.3 &	$\pm$ & 40.4\\
2.35 & 0.404 & 0.80 & 165.5 &	$\pm$ & 14.6 &	$\pm$ & 19.4 & -26.8 &	$\pm$ & 15.1 &	$\pm$ & 16.1 & 6.5 &	$\pm$ & 27.5 &	$\pm$ & 16.3\\
2.35 & 0.404 & 1.25 & 114.4 &	$\pm$ & 12.1 &	$\pm$ & 20.4 & -9.7 &	$\pm$ & 12.9 &	$\pm$ & 17.9 & -29.9 &	$\pm$ & 21.1 &	$\pm$ & 24.1\\
2.35 & 0.404 & 1.75 & 84.0 &	$\pm$ & 24.7 &	$\pm$ & 55.2 & 1.4 &	$\pm$ & 27.9 &	$\pm$ & 76.6 & -12.0 &	$\pm$ & 38.4 &	$\pm$ & 100.8\\
2.73 & 0.343 & 0.35 & 94.2 &	$\pm$ & 20.7 &	$\pm$ & 14.9 & -28.5 &	$\pm$ & 29.4 &	$\pm$ & 16.0 & 46.0 &	$\pm$ & 48.7 &	$\pm$ & 29.3\\
2.73 & 0.343 & 0.50 & 79.1 &	$\pm$ & 6.1 &	$\pm$ & 3.2 & -3.8 &	$\pm$ & 8.3 &	$\pm$ & 6.9 & 18.8 &	$\pm$ & 19.3 &	$\pm$ & 15.1\\
2.73 & 0.343 & 0.80 & 58.9 &	$\pm$ & 3.4 &	$\pm$ & 2.3 & 12.5 &	$\pm$ & 4.3 &	$\pm$ & 4.4 & -8.5 &	$\pm$ & 10.7 &	$\pm$ & 5.5\\
2.73 & 0.343 & 1.25 & 28.6 &	$\pm$ & 2.4 &	$\pm$ & 2.9 & -0.2 &	$\pm$ & 3.2 &	$\pm$ & 1.2 & -4.2 &	$\pm$ & 7.2 &	$\pm$ & 9.8\\
2.73 & 0.343 & 1.75 & 18.7 &	$\pm$ & 2.2 &	$\pm$ & 2.7 & -4.8 &	$\pm$ & 3.0 &	$\pm$ & 2.4 & 2.5 &	$\pm$ & 6.0 &	$\pm$ & 9.8\\
2.77 & 0.424 & 0.50 & 164.4 &	$\pm$ & 20.7 &	$\pm$ & 21.0 & -53.5 &	$\pm$ & 23.4 &	$\pm$ & 25.3 & 26.9 &	$\pm$ & 36.6 &	$\pm$ & 33.4\\
2.77 & 0.424 & 0.80 & 100.9 &	$\pm$ & 7.5 &	$\pm$ & 11.5 & 12.2 &	$\pm$ & 8.4 &	$\pm$ & 13.3 & -17.2 &	$\pm$ & 16.9 &	$\pm$ & 22.4\\
2.77 & 0.424 & 1.25 & 67.8 &	$\pm$ & 5.5 &	$\pm$ & 7.4 & 7.9 &	$\pm$ & 6.4 &	$\pm$ & 6.1 & -29.8 &	$\pm$ & 12.6 &	$\pm$ & 13.7\\
2.77 & 0.424 & 1.75 & 45.3 &	$\pm$ & 6.3 &	$\pm$ & 6.9 & -4.4 &	$\pm$ & 7.6 &	$\pm$ & 10.3 & 9.2 &	$\pm$ & 11.8 &	$\pm$ & 17.6\\
3.25 & 0.430 & 0.50 & 108.4 &	$\pm$ & 20.7 &	$\pm$ & 14.8 & -22.2 &	$\pm$ & 27.1 &	$\pm$ & 17.5 & 21.1 &	$\pm$ & 42.7 &	$\pm$ & 23.3\\
3.25 & 0.431 & 0.80 & 62.2 &	$\pm$ & 5.3 &	$\pm$ & 4.7 & 9.8 &	$\pm$ & 7.0 &	$\pm$ & 4.7 & -23.3 &	$\pm$ & 14.8 &	$\pm$ & 11.9\\
3.25 & 0.431 & 1.25 & 47.1 &	$\pm$ & 4.2 &	$\pm$ & 3.9 & -3.6 &	$\pm$ & 5.5 &	$\pm$ & 8.6 & -0.6 &	$\pm$ & 11.8 &	$\pm$ & 136.3\\
3.25 & 0.431 & 1.75 & 30.6 &	$\pm$ & 4.9 &	$\pm$ & 3.5 & -7.3 &	$\pm$ & 6.9 &	$\pm$ & 4.5 & 6.3 &	$\pm$ & 11.7 &	$\pm$ & 13.2\\
3.30 & 0.497 & 1.75 & 128.6 &	$\pm$ & 38.4 &	$\pm$ & 35.0 & -6.8 &	$\pm$ & 42.0 &	$\pm$ & 19.6 & 17.4 &	$\pm$ & 77.0 &	$\pm$ & 52.1\\
3.69 & 0.451 & 0.80 & 68.1 &	$\pm$ & 11.7 &	$\pm$ & 5.9 & -12.1 &	$\pm$ & 18.2 &	$\pm$ & 5.5 & 6.9 &	$\pm$ & 47.2 &	$\pm$ & 25.2\\
3.77 & 0.513 & 0.80 & 71.4 &	$\pm$ & 43.1 &	$\pm$ & 10.8 & 15.2 &	$\pm$ & 57.8 &	$\pm$ & 25.4 & -38.8 &	$\pm$ & 76.2 &	$\pm$ & 30.0\\
3.77 & 0.514 & 1.25 & 56.5 &	$\pm$ & 14.3 &	$\pm$ & 7.3 & 11.5 &	$\pm$ & 20.2 &	$\pm$ & 11.1 & -29.6 &	$\pm$ & 34.9 &	$\pm$ & 22.9\\
3.77 & 0.513 & 1.75 & 57.2 &	$\pm$ & 17.6 &	$\pm$ & 9.1 & -3.4 &	$\pm$ & 23.9 &	$\pm$ & 8.8 & -17.4 &	$\pm$ & 34.3 &	$\pm$ & 16.0\\
4.24 & 0.540 & 1.25 & 100.7 &	$\pm$ & 30.2 &	$\pm$ & 12.7 & -46.3 &	$\pm$ & 44.9 &	$\pm$ & 15.4 & 48.5 &	$\pm$ & 72.4 &	$\pm$ & 20.6\\
\end{longtable}

\end{widetext}

\bibliography{eta_paper}

\end{document}